\documentclass[aps,prx,twocolumn,superscriptaddress,longbibliography]{revtex4-2}
\usepackage{amsmath, amsthm}
\usepackage{keyval}
\usepackage{graphicx,latexsym}
\usepackage{dcolumn}
\usepackage{bm}
\usepackage{color}
\usepackage{ulem} 
\usepackage{hyperref}
\usepackage{xr}

\newcommand{\avg}[1]{\langle #1 \rangle}

\definecolor{magenta}{rgb}{1,0,1}
\usepackage{color}

\hypersetup{
    colorlinks=true, 
    linktoc=all,   
    linkcolor=blue,  
}

\begin{document}
\title{First principles prediction of structural distortions in the cuprates and their impact on electronic structure}
\author{Zheting Jin}
\affiliation{Department of Applied Physics, Yale University, New Haven, Connecticut 06520, USA}
\author{Sohrab Ismail-Beigi}
\affiliation{Department of Applied Physics, Yale University, New Haven, Connecticut 06520, USA}
\affiliation{Department of Physics, Yale University, New Haven, Connecticut 06520, USA}
\affiliation{Department of Mechanical Engineering and Materials Science, Yale University, New Haven, Connecticut 06520, USA}
\date{\today}
\begin{abstract}
Materials-realistic microscopic theoretical descriptions of copper-based superconductors are challenging due to their complex crystal structures combined with strong electron interactions. Here, we demonstrate how density functional theory can accurately describe key structural, electronic, and magnetic properties of the normal state of the prototypical cuprate Bi$_2$Sr$_2$CaCu$_2$O$_{8+x}$ (Bi-2212).  We emphasize the importance of accounting for energy-lowering structural distortions, which then allows us to: (a) accurately describe the insulating antiferromagnetic (AFM) ground state of the undoped parent compound (in contrast to the metallic state predicted by previous {\it ab initio} studies); (b) identify numerous low-energy competing spin and charge stripe orders in the hole-overdoped material nearly degenerate in energy with the AFM ordered state, indicating strong spin fluctuations; (c) predict the lowest-energy hole-doped crystal structure including its long-range structural distortions and oxygen dopant positions that match high-resolution scanning transmission electron microscopy (STEM) measurements; and (d) describe electronic bands near the Fermi energy with flat antinodal dispersions and Fermi surfaces that in agreement with angle-resolved photoemission spectroscopy (ARPES) measurements and provide a clear explanation for the structural origins of the so-called ``shadow bands''. We also show how one must go beyond band theory and include fully dynamic spin fluctuations via a many-body approach when aiming to make quantitative predictions to measure the ARPES spectra in the overdoped material.  Finally, regarding spatial inhomogeneity, we show that the local structure at the CuO$_2$ layer, rather than dopant electrostatic effects, modulates the local charge-transfer gaps, local correlation strengths, and by extension the local superconducting gaps.
\end{abstract}
\maketitle
\section{Introduction}

The cuprate superconductors continue to be a fascinating and actively researched class of materials.  In addition to their superconducting phases, their normal state has attracted broad research interest due to a wide range of unusual properties. 
Understanding the physical origin of the AFM insulating phase \cite{kastner1998magnetic, ho2001nature}, the pseudogap \cite{warren1989cu, homes1993optical, loeser1996excitation, ding1996spectroscopic}, the flat antinodal dispersion \cite{andersen1994plane, liechtenstein1996quasiparticle, he2021superconducting, li2021superconductor}, the strange metallicity \cite{gurvitch1987resistivity, martin1990normal, cooper2009anomalous, greene2020strange, ayres2021incoherent}, and the presence of quantum critical fluctuations \cite{dai2001evolution, le2011intense, dean2013persistence, keimer2015quantum} in the normal state can provide important insights into the underlying mechanisms that can give rise to the superconductivity.   
Despite plenty of proposed possible mechanisms, including competing orders \cite{kivelson1998electronic, lee2006doping, chang2012direct} and preformed pairs  \cite{emery1995importance, corson1999vanishing, li2010diamagnetism, dubroka2011evidence}, the physical origin of this complicated normal state is still unclear. 

A comprehensive description of the normal state is challenging due to the strong electronic interactions combined with the structural complexity of typical doped cuprates.  
While the effect of strong electronic interactions has been extensively studied using accurate many-body methods such as density matrix renormalization group (DMRG) \cite{white1992density, schollwock2005density} or quantum Monte Carlo (QMC) \cite{ceperley1986quantum, foulkes2001quantum}, these studies are often based on idealized effective model Hamiltonian, where the hopping strengths are averaged and symmetrized for simplicity \cite{cui2020ground, white2015doping, huang2017numerical}.  However, realistic structural distortions in cuprates can result in significant changes to the materials such as additional symmetry breaking \cite{gao1988incommensurate, mans2006experimental}.  In particular, the structural distortion in cuprates can greatly modify the superconducting gap \cite{mcelroy2005atomic, andersen2007superconducting} and local pairing interactions \cite{foyevtsova2009effect, gastiasoro2018enhancing, nunner2005dopant, maska2007inhomogeneity, khaliullin2010enhanced, litak2009charge, romer2013modulations, johnston2009impact}.   These properties are missing in the symmetrized effective Hamiltonians.  The inclusions of complex lattice distortions of the native material or as introduced by dopants and impurities necessitate a realistic and detailed understanding of the materials from first principles.  

Density functional theory \cite{hohenberg1964inhomogeneous, kohn1965self} (DFT) offers a potent foundational method for investigating the ground-state properties of materials from first principles.  For cuprates, DFT has played a pivotal role in constructing effective model Hamiltonians \cite{hybertsen1989calculation, martin1996electronic, markiewicz2005one, korshunov2005hybrid, delannoy2009low, hanke20103, hirayama2019effective, ohgoe2020ab, nilsson2019dynamically, cui2022systematic, cui2023ab, moree2022ab, schmid2023superconductivity}.  These works studied the high-symmetry cuprate crystal structure using DFT and extracted the low-energy effective model for further quantum many-body studies.   However, to ensure the accuracy and realism of these models, it is essential that the DFT calculations capture correctly both the structural properties and the predominant electronic properties of the ground state.  Recent DFT studies on transition metal oxides have highlighted the significance of allowing energy-lowering structural distortions to achieve high-quality predictions of materials properties, such as band gaps at eV scale \cite{varignon2019origin, wang2021mass,zunger_bridging_2022}.  

Bismuth strontium calcium copper oxide Bi$_2$Sr$_2$CaCu$_2$O$_{8+x}$ (BSCCO or Bi-2212) \cite{maeda1988new} is one of the most intensively studied cuprates and is the focus of in this work.  However, prior DFT studies faced challenges due to its intricate structural distortions and superlattice modulations \cite{petricek1990x, levin1994causes, slezak2008imaging, he2008supermodulation}.  We demonstrate that by employing modern DFT exchange-correlation functionals and providing an accurate description of the crystalline structure including energy-lowering lattice distortions, we can directly describe the antiferromagnetic insulating ground state of undoped Bi-2212 ($x=0$), the experimentally-observed crystal structure of hole-doped Bi-2212 ($x\approx0.25$), the presence of competing magnetic and charge stripe orders, as well as crucial details of photoemission spectra such as ``shadow bands''.  
This means that based on DFT calculations, we can correctly ascribe certain experimental observations to specific structural motifs (e.g., shadow bands) while simultaneously helping build microscopically well-justified model Hamiltonians that allow us to study the effects of strong electron correlations further.  Our many-body calculations based on such models predict spatial modulations of charge-transfer gaps and correlation strength, consistent with experimental observations \cite{o2022electron, mcelroy2005atomic, andersen2007superconducting}.  Interestingly, structural modulation of the CuO$_2$ plane is the dominant cause of this phenomenon, far more important than the electrostatic effects from dopant atoms that were assumed to be important in earlier model calculations  \cite{foyevtsova2009effect, gastiasoro2018enhancing, maska2007inhomogeneity, khaliullin2010enhanced, johnston2009impact}. 

\section{Undoped system}
\label{sec:undoped}
\begin{figure}[t]
\begin{center}
\includegraphics[scale=0.4]{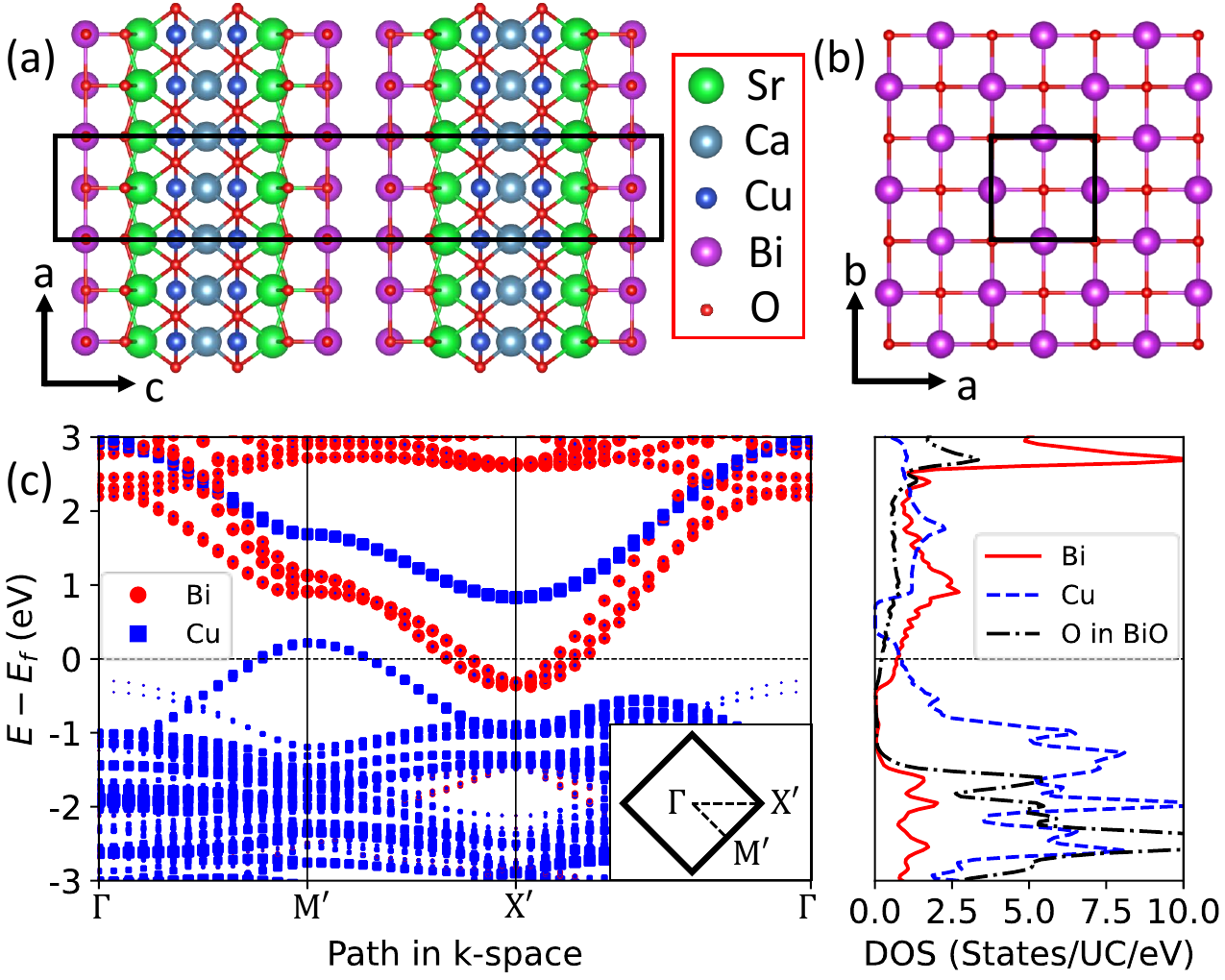}
\end{center}
\caption{
Crystal and electronic structure of high-symmetry undoped Bi-2212.  (a) Side view of the crystal structure along the $b$-axis.  Sr, Ca, Cu, Bi, and O atoms are marked by green, gray, blue, purple, and red balls, respectively.  The black square in the crystal structure marks the G-AFM unit cell. 
(b) Top view of the BiO layer from the $c$-axis.  
(c) Projected band structures (left) and density of states (DOS) (right) from DFT calculation, where the Fermi energy is set to be the reference energy.  The inset of the band structures shows the first Brillouin zones (BZ) of the 60-atom supercell.  The unit of DOS is the number of states per unit cell (UC) per eV.  Red circles and solid lines show Bi $p$-orbitals; blue squares and dash lines show Cu $d$-orbitals; black dash-dotted line shows O $p$-orbitals in BiO layers.
}
\label{fig:high_symmetry}
\end{figure}
The undoped Bi-2212 possesses a bilayer crystal structure as depicted in Fig. \ref{fig:high_symmetry}(a)(b), where each bilayer comprises two CuO layers separated by one Ca layer and sandwiched between SrO and BiO layers.  The smallest unit cell contains 30 atoms and crystallizes in the tetragonal I4/mmm space group, with only one Cu atom in each CuO layer \cite{liang1988crystal, 10.1063/1.4812323,materialproject}.  However, this small unit cell leads to a false non-magnetic ground state due to the artificial assumption of translational invariance of the Cu local moments \cite{yelpo2021electronic}.  In addition, the tetragonal space group of this small unit cell is inconsistent with the orthorhombicity observed in experiments \cite{zeljkovic2012scanning}.  Here, we start with a larger supercell (60 atoms/cell) with two Cu atoms in each CuO layer, which allows for structural orthorhombicity and the spontaneous symmetry breaking of the magnetic local moments.  We ignore the superlattice modulation distortion for now and focus on the distortions in BiO layers.  As we will show in the following, the distortions in BiO layers are crucial to raising the antibonding Bi-O band to higher energy.

For simplicity, we first compare different distortion patterns in bulk calculations.  After finding the most energetically favorable distortions, we then perform slab calculation to compare with available undoped Bi-2212 experiments such as STM and thin film transports.  Because the inter-bilayer van der Waals interactions in Bi-2212 are weak, the slab calculations show very small differences with the bulk results.  This is also the reason why many prior studies only focused on slab calculations \cite{wang2005ab, he2006local, foyevtsova2010modulation}. 

First, we perform a relaxation starting with a crystal structure drawn for available databases \cite{materialproject, 10.1063/1.4812323, subramanian1988new} to identify the nearest local minimum with the same symmetry, referred to as the ``high-symmetry'' structure.  We find a G-AFM ordered ground state with local Cu magnetic moments of $\pm 0.45\mu_B$, while the non-magnetic state is about 0.25 eV/Cu higher in energy.  These local moments agree with the experimental measurements, typically falling within the range of 0.4-0.6$\mu_B$ in cuprates without chlorine  \cite{schrieffer2007handbook}. Fig. \ref{fig:high_symmetry}(c) shows the projected band structure and density of states of this AFM ground state which is metallic.  The AFM order opens a gap of about 0.6 eV for Cu $d$-orbitals between $M'$ and $X'$.  The Bi-O in-plane coupling opens an about 1.0 eV bonding gap below the Fermi level.  As we will discuss in the following, this high-symmetry crystal is not the most energetically favorable structure for undoped BSCCO. 


\begin{figure}[t]
\begin{center}
\includegraphics[scale=0.4]{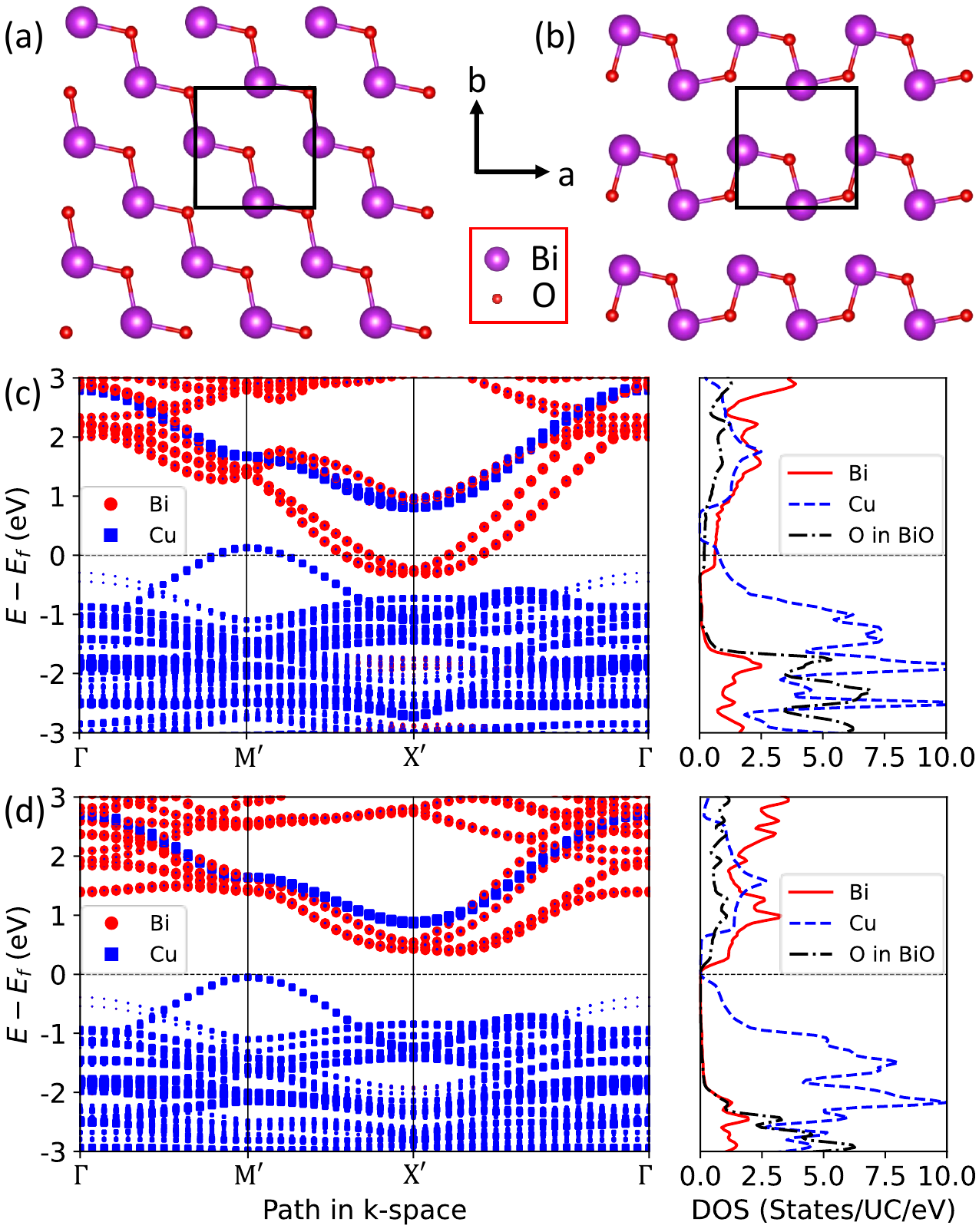}
\end{center}
\caption{
Crystal and electronic structure of two stable low-symmetry crystals of undoped Bi-2212.  
(a)(b) The top views of the ``zigzag'' and ``orthorhombic'' distortion patterns of the BiO layer, where the orthorhombic distortion pattern in (b) is the most energetically favorable structure.  Large purple and small red balls represent Bi and O atoms.  The black squares in the crystal structure illustrate the G-AFM unit cells. 
(c)(d) Projected band structures (left) and the density of states (right) of the crystal structures in (a) and (b), separately.  Red circles and solid lines show Bi $p$-orbitals; blue squares and dash lines show Cu $d$-orbitals; black dash-dotted line shows O $p$-orbitals in BiO layers.
}
\label{fig:low_symmetry}
\end{figure}

To improve the theoretical description, allowing crystal structural distortions in the BiO layers can further lower the total energy of the calculation.  Using the conjugate gradient algorithm for structural relaxation, we optimize the structures by initially lowering the symmetry in the BiO layers manually.  Fig. \ref{fig:low_symmetry}(a)(b) depicts two typical stable or meta-stable BiO layer patterns whose ground states are both G-AFM ordered on the CuO$_2$ planes.  These two distortions have been studied separately in prior studies \cite{bellini2004structure, lin2006raising, nokelainen2020ab}.  Here we will compare the effects of different distortion patterns on electronic structures and demonstrate how the distortions help raise the BiO antibonding band.   

Fig. \ref{fig:low_symmetry}(a) shows a zigzag pattern, where all the oxygen atoms on BiO layers are moved along the diagonal direction of the in-plane unit cell by the same amount.  With all BiO layers displaying this zigzag distortion pattern, the AFM ground state energy is about 0.29 eV/Cu lower than the one with the high-symmetry structure in Fig. \ref{fig:high_symmetry}(b), and the local Cu magnetic moments are $\pm 0.48\mu_B$, slightly larger than the high-symmetry case.  The BiO bonding gap increases to about 1.5 eV.  

Fig. \ref{fig:low_symmetry}(b) shows an orthorhombic pattern, characterized by the lowest symmetry among all three structures presented in Fig. \ref{fig:high_symmetry} and \ref{fig:low_symmetry}.  This pattern further breaks the mirror symmetry of the crystal, resulting in an orthorhombic lattice consistent with observations in experiments \cite{zeljkovic2012scanning}.  This orthorhombic pattern exhibits the lowest AFM ground state energy, 0.5 eV/Cu lower than the high-symmetry case.  It also features the largest local Cu magnetic moments of $\pm 0.53\mu_B$ as well as the largest BiO bonding gap size of about 1.9 eV among all three Bi-O structural motifs mentioned above.  

The electronic structures of the distorted crystals in Figs. \ref{fig:low_symmetry}(a) and (b) are presented in Figs. \ref{fig:low_symmetry}(c) and (d), respectively.  Compared to the electronic structure of the high-symmetry case in Fig. \ref{fig:high_symmetry}(c), most of the changes are observed within the Bi bands due to the BiO distortion patterns.  Although the zigzag distortion pattern helps reduce the size of the Bi electron pocket, it's only the lowest-energy orthorhombic distortion pattern that elevates the entire Bi bands above the Fermi level, resulting in an insulating AFM ground state.   

A straightforward microscopic picture helps elucidate how the distortions contribute to elevating the Bi bands to higher energy levels.  The densities of states (DOS) plots reveal that the in-plane BiO system possesses filled low-energy bonding states dominated by oxygen and antibonding states dominated by bismuth, and the bonding/anti-bonding gaps are centered at about an energy 1 eV below the Fermi energy for all three structures.  However, the size of the bonding/anti-bonding gap varies with the distortion pattern.  The DOS shows that this gap is smallest for the high-symmetry structure in Fig. \ref{fig:high_symmetry}(c) ($\sim$1.0 eV) and largest for the orthorhombic structure in Fig. \ref{fig:low_symmetry}(d) ($\sim$1.9 eV).  The gap size difference arises from different Bi-O hybridization strengths among the three structures.  The coupling is strong enough only in the orthorhombic structure to lift the anti-bonding state above the Fermi level, while the couplings in high-symmetry and zigzag structures are too weak, resulting in metallic states. 

\begin{table}
\begin{tabular}{c|ccc}
Structure &\ hopping (eV) & \ bond length (\AA) & \ bond angle ($^{\circ}$) \\
\hline
1 & 1.0 & 2.65 & 90 \\
2 & 1.8 & 2.28 & 112 \\
3 & 2.3 & 2.17 & 93 \\
4 & 0.9 & 2.65 & 93 \\
\end{tabular}
\caption{Bi-O hopping strengths, Bi-O bond lengths, and O-Bi-O bond angles in the BiO plane bond for different distorted structures:  structure 1 is the high-symmetry structure of Fig. \ref{fig:high_symmetry}(b); 2 is the zigzag distorted structure of Fig. \ref{fig:low_symmetry}(a);  3 is the orthorhombic distorted structure of Fig. \ref{fig:low_symmetry}(b); 4 has the same geometry as structure 3 but with stretched in-plane lattice constant to match the Bi-O bond length of structure 1.}
\label{tab:hop_strength}
\end{table}

This microscopic picture is consistent with a structural analysis.  Generally, hopping strengths depend directly on local structural properties such as bond lengths and angles.  We compare four different structures in Table \ref{tab:hop_strength}: 1. the high-symmetry structure from Fig. \ref{fig:high_symmetry}; 2. the zigzag distorted structure from Fig. \ref{fig:low_symmetry}(a);  3. the orthorhombic distorted structure from Fig. \ref{fig:low_symmetry}(b); 4. the orthorhombic distorted structure with expanded in-plane lattice constant.  Structures 1-3 are stable/meta-stable relaxed structures predicted by DFT, and structure 4 is prepared to have the same distortion pattern as structure 3 but the same bond length as the high-symmetry structure 1.  To extract the hopping strength, we compute the tight-binding Kohn-Sham Hamiltonian on the maximally localized Wannier basis \cite{marzari1997maximally} extracted from our DFT calculations using Wannier90 \cite{pizzi2020wannier90}.  The projected Wannier orbitals encompass the $p$-orbitals of Bi and O, as well as the $d$-orbitals of Cu, which are sufficient to describe the bands near the Fermi level \cite{supp}.  Table \ref{tab:hop_strength} lists the Bi-O hopping (tunneling) matrix element for nearest-neighbor in-plane Bi-O pairs for different structures.  Upon comparing these four structures, we observe that the bond length has the most significant impact on the hopping strength (shorter bonds give larger hoppings, as expected).  The bond angles do influence the hopping strengths but only modestly (compare structures 1 and 4).  Notably, the hopping strength in structure 3 (the lowest energy structure predicted by DFT) is the only one large enough to open a band gap by raising the Bi electron pockets (originating from the anti-bonding BiO bands) above the Fermi level, and this leads to the insulating ground state.

\begin{figure}[t]
\begin{center}
\includegraphics[scale=0.34]{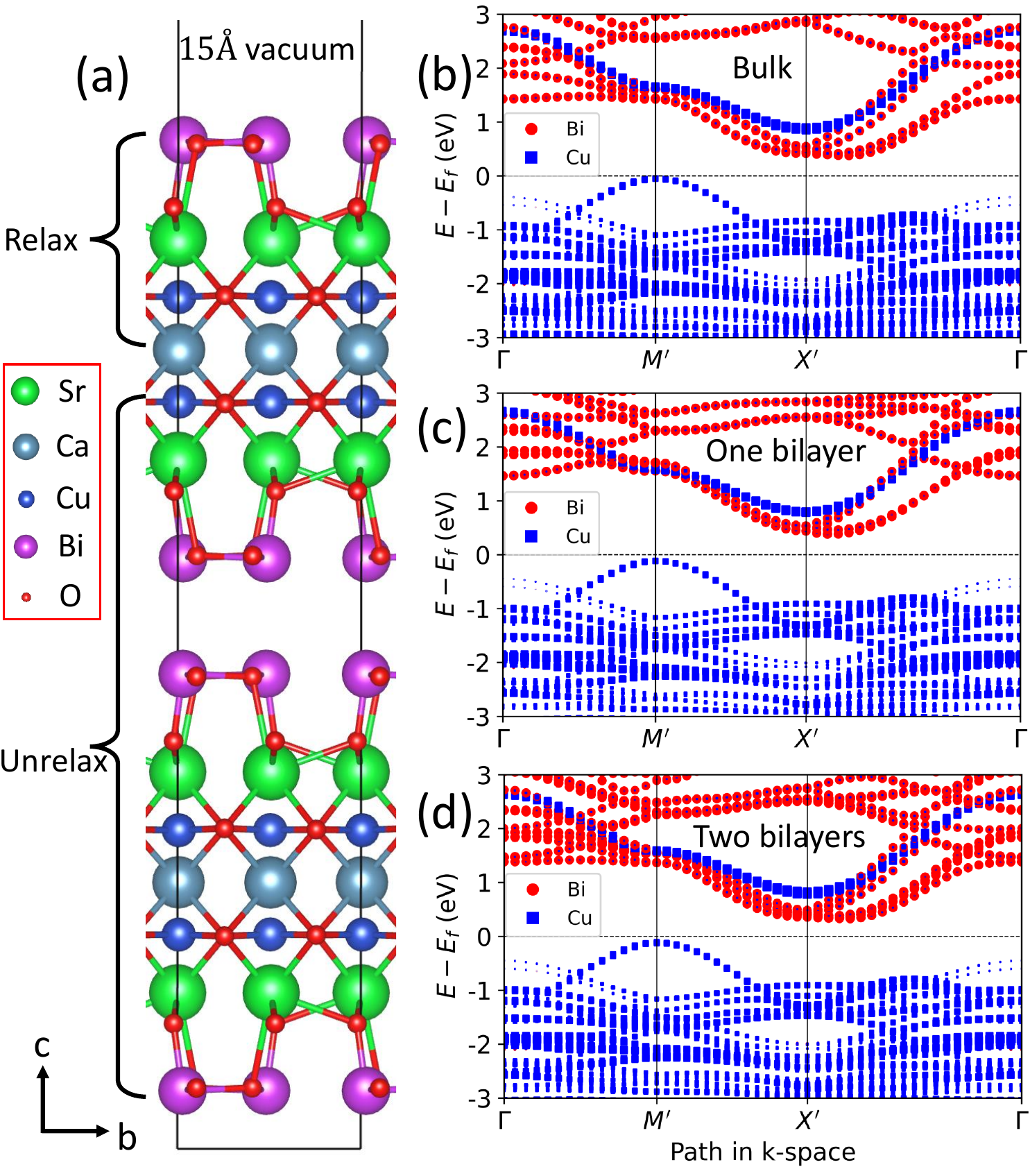}
\end{center}
\caption{
(a) Illustration of the slab geometry where 15 \AA vacuum is added between periodic copies of the slabs along $c$.  Several surface atomic layers including BiO, SrO, CuO, and Ca layers are relaxed to study the surface effect.  The remaining atoms are frozen in their bulk configurations.  (b) The projected band structure of bulk BSCCO calculation adapted from Fig. \ref{fig:low_symmetry}(d).  Red circles and blue squares represent Bi and Cu projections.  (c) The projected band structure of a slab calculation comprised of only the upper bilayer in (a).  (d) The projected band structure of the full unit cell containing two bilayers as shown in (a).
}
\label{fig:app_surface}
\end{figure}

Our last set of calculations for undoped materials will describe thin films by using slab calculations.  We start with the most energetically favorable bulk structure and insert 15 \AA{} of vacuum between BiO layers to create surfaces and slabs. Fig. \ref{fig:app_surface}(a) illustrates the structural model: the four atomic layers closest to the slab's surface are relaxed to allow for possible surface reconstructions. Fig. \ref{fig:app_surface}(b)-(d) show the projected band structures of bulk, a slab of one bilayer, and a slab of two bilayers.  They all have very similar band structures near the Fermi level.  There are some very modest surface effects on the unoccupied Bi-derived bands about 1 eV above the Fermi level, while the Cu-derived bands, both occupied and empty, are not affected by the surfaces.  None of this is surprising: the Cu layers are a few atomic layers away from the surface where the Bi layer resides.  All three calculations find insulating band structures.  

We conclude this section by comparing our results with experiments and prior theories.  Experimentally, bulk Bi-2212 crystals are conducting due to hole hoping from inevitable excess oxygen $(x>0)$, so comparing our undoped calculations to experiments on the bulk is not fruitful.  However, undoped thin films of Bi-2212 ($x=0$) have been realized experimentally: scanning tunneling microscopy (STM) \cite{ruan2016relationship, kowalski2021oxygen, wang2023correlating} and transport \cite{yu2019high, wang2021superconductor} experiments have found an AFM insulating ground state in undoped thin films Bi-2212.  (About $x\approx 6\%$ hole-doping is required to turn the insulating Bi-2212 film into a superconductor \cite{yu2019high, wang2021superconductor}.)  Theoretically, some aspects of the undoped ground state have not been described well by prior DFT work which may have hampered further theoretical studies of the effects of doping and other perturbations on the system at a microscopic level.  For example, prior DFT works were unable to simultaneously reproduce the AFM order and the insulating behavior of undoped Bi-2212 whether modeling the bulk \cite{yelpo2021electronic, lin2006raising, nokelainen2020ab}, thin film or or slabs \cite{wang2005ab, he2006local, foyevtsova2010modulation}, or when using larger more realistic unit cells that include the observed crystal modulation  \cite{he2008supermodulation, fan2011modulation}.  The prior work found undoped Bi-2212 to be metallic and ``self-doped'' due to Bi electron pockets at the Fermi surface (pockets that have not been seen in experiments).  Some prior works \cite{lin2006raising, he2006local, foyevtsova2010modulation} resorted to additional ``manual'' hole doping to eliminate the finite Bi density of states: while this is a sensible workaround, it would be preferable to have a calculation of the undoped material reflect the insulating experimental system.  

Our calculations provide a simple and correct description of the undoped material as an AFM insulator, in agreement with available experiments.  The reason for success is straightforward and combines two aspects of prior works in a single calculation: more capable exchange-correlation functions (such as GGA+U or SCAN+U) must be combined with thorough relaxation of the microscopic structure to allow for energy-lowering structural distortions to occur and open up the band gap.

\section{Oxygen-doped system}

\begin{figure}[t]
\begin{center}
\includegraphics[scale=0.4]{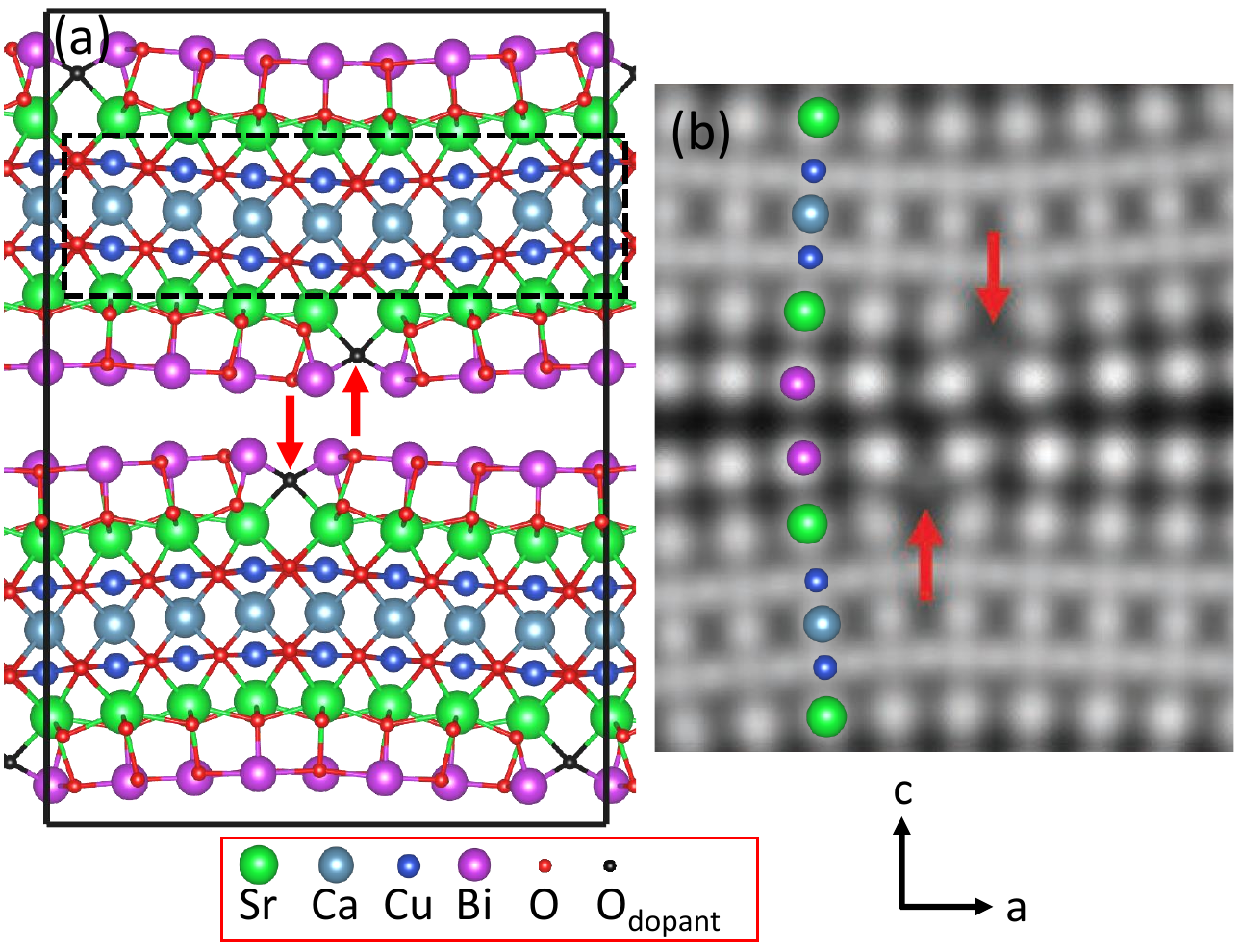}
\end{center}
\caption{
(a) DFT predicted crystal structure of Bi$_2$Sr$_2$CaCu$_2$O$_{8.25}$, where the hole doping level is $x=0.25$.  Sr, Ca, Cu, Bi, O, and dopant O atoms are marked by green, gray, blue, purple, red, and black balls, respectively.  Red arrows further highlight the O dopants.  The black solid square in the crystal structure marks the 244-atom unit cell of the doped crystal.  The black dashed rectangle highlights the CuO layers in one of the bilayers in the unit cell.  (b) Measured crystal structure of BSCCO with hole doping of $x\approx 22\%$ using STEM adapted from Ref. \cite{song2019visualization}.  The red arrows are from the original work, pointing out the oxygen dopants.  We show a column of colorful balls for easy identification of the atomic species.  
}
\label{fig:244_crystal}
\end{figure}

Commencing with undoped Bi-2212, hole doping is incorporated through interstitial oxygen dopants within the material. These additional hole dopants give rise to a diverse range of physical properties including superconductivity, the pseudogap phenomenon, and the presence of Fermi surface. The subsections below present our findings on various aspects of the hole-doped system at 25\% (overdoped): the hole-doped crystal structure including the long-range superlattice modulation and associated placement of the oxygen dopants; the nature of the low-energy magnetic states including spin and charge stripes; and a detailed analysis of the ARPES spectra at the Fermi level including the effects of structural distortions in creating the shadow bands as well as the effect of many-body effects and fluctuating magnetic orders on these spectra.

\subsection{Crystal structure}
For the oxygen-doped Bi$_2$Sr$_2$CaCu$2$O$_{8+x}$ system with $x=0.25$, we explicitly include the  superlattice modulation \cite{gao1988incommensurate, petricek1990x, levin1994causes, slezak2008imaging, he2008supermodulation}.  Experimentally, this modulation exists with a period between 4-5 unit cells regardless of doping level \cite{song2019visualization, levin1994causes, le1989origin, grebille1996static}, and theoretically confirmed that the modulation is an intrinsic behavior of BSCCO in that it occurs even without doping \cite{he2008supermodulation}.  Of course, in a computational model using periodic boundary conditions, the superlattice modulation must conform to the size of the computational supercell, e.g., 5 unit cells in Ref.~\cite{he2008supermodulation} and 4 unit cells in our work below; either choice is reasonable given the range of 4-5 unit cells found in experiments.  We choose 4 unit cells for purely pragmatic reasons: the ARPES experiments \cite{he2021superconducting} we will compare to below (concerning the shadow bands) are estimated to have a doping of $0.22<x<0.24$ (based on the superconducting temperature) or $x=0.26$ (based on the Fermi surface volume).  Our choice of $x=0.25$ is suitable.  Finally, unlike a calculation with periodic boundary conditions, the positions of the oxygen dopants may be disordered in the real material which will modify spectroscopic intensities.  However, given the above-described robustness of the superlattice modulation with respect to doping, we believe that the superlattice modulation (and spectroscopic features deriving from it such as the shadow bands) should only suffer modest broadening and generally remain intact with respect to oxygen disorder.

For the chosen supercell which accommodates a period-4 superlattice, we explore various possible positions for the oxygen dopants.  These structural properties are essential for describing the BiO layers properly, and they also have a significant impact on the superconducting gap \cite{andersen2007superconducting, massee2020atomic}. 

Prior work using smaller unit cells has shown that the lowest-energy oxygen dopant positions are between the Bi and Sr layers \cite{he2006local, nokelainen2020ab}.  Hence we follow the prior work to look for optimized oxygen dopant positions around the Bi and Sr layers.  We have tested multiple oxygen dopant positions, and Fig. \ref{fig:244_crystal} (a) shows our optimized lowest-energy crystal structure for Bi$_2$Sr$_2$CaCu$_2$O$_{8.25}$, where four oxygen dopants are added into a stoichiometric undoped Bi-2212 unit cell of 240 atoms which is a  $4\times1\times1$ enlargement of the 60-atom unit cell of Sec. \ref{sec:undoped}.  Several metastable crystals are listed in the supplementary material \cite{supp}, showing an energy cost of at least 900 meV/dopant to move oxygen dopants between two BiO layers, and at least 168 meV/dopant to move dopants between BiO and SrO layers.  Regardless of the dopant positions, the relaxed superlattice modulation remains in the same pattern, consistent with the modulations found in prior experimental and theoretical works \cite{gao1988incommensurate, petricek1990x, levin1994causes, slezak2008imaging, he2008supermodulation}.

We conclude that the most stable positions for the oxygen dopants are located at the necking region between BiO and SrO layers as highlighted in Fig.~\ref{fig:244_crystal}.  This position agrees with the in-plane dopant coordinates determined by STM experiments \cite{mcelroy2005atomic}, but STM  is performed on the surface and cannot describe the dopant positions along the out-of-plane direction.  Hence, we compare directly to recent high-resolution STEM measurements \cite{song2019visualization} which can provide orthogonal information on atomic positions: we see excellent agreement with the STEM-observed structure as shown in Fig. \ref{fig:244_crystal} (b).  Hence, we are confident in our predicted ground-state structure for the doped crystal and use it below to predict a variety of low-energy magnetic and charge orderings.

\subsection{Magnetic and charge ordering}
\label{sec:244_magnetic_charge}

With the lattice structure of our doped BSCCO confirmed, we turn our attention to the electronic spin and charge distribution within this system. We will show that many spatially spin- and/or charge-ordered electronic states are almost degenerate in energy with the G-type (checkerboard) AFM-ordered state.  This indicates the presence of strong spin and charge fluctuations and competing orders in this overdoped normal state, and such fluctuations are thought to be a possible physical origin of the pseudogap phase in the cuprates \cite{zhang2020competing}.

\begin{figure}[t]
\begin{center}
\includegraphics[scale=0.45]{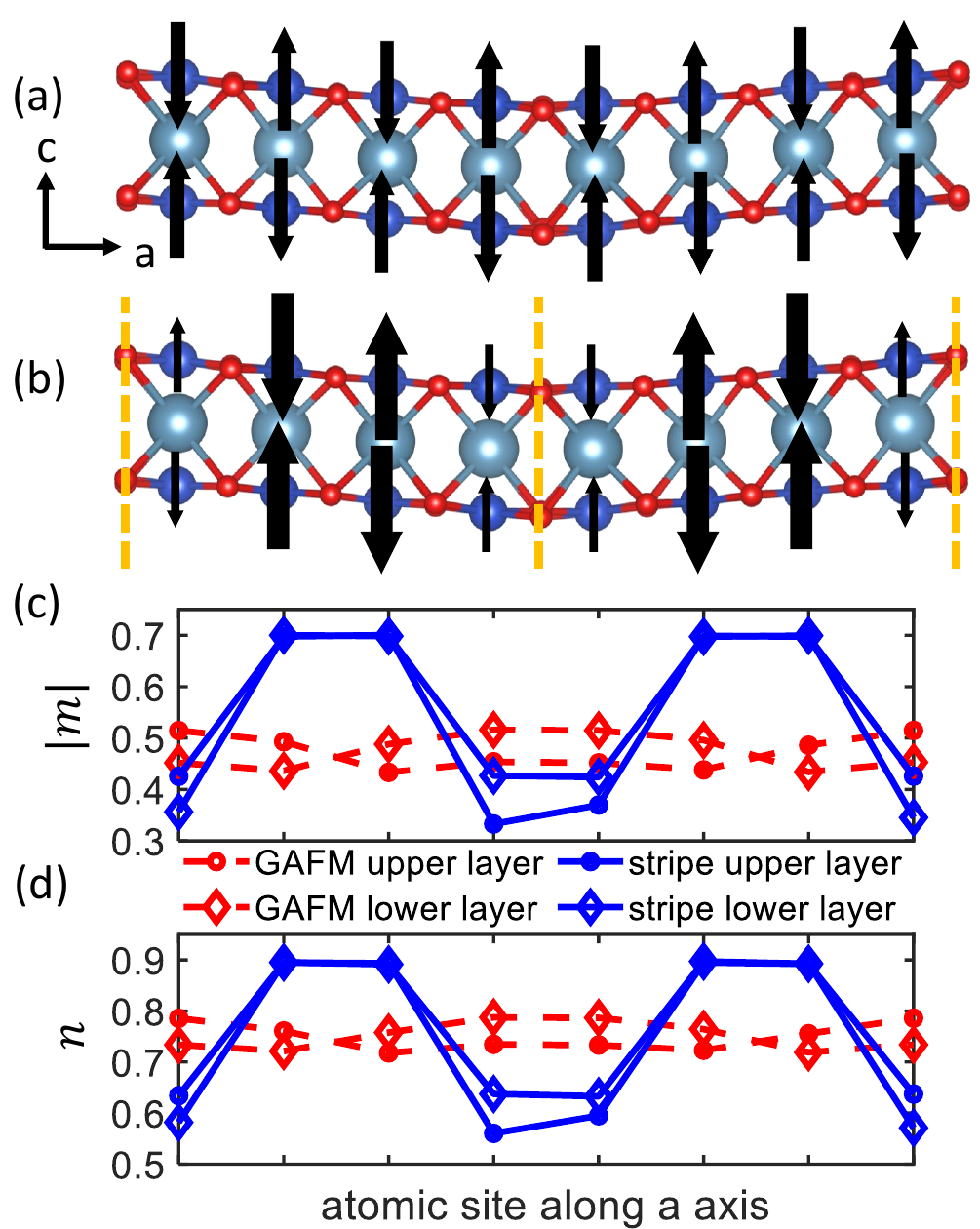}
\end{center}
\caption{
Competing G-AFM and stripe order phases in a CuO bilayer.  (a,b) Illustrations of G-AFM (a) and a typical stripe order phase (b) for the bilayer in the black dashed square of Fig. \ref{fig:244_crystal}.  The black arrows represent the local moments on Cu atoms, with exaggerated length and thickness highlighting their relative magnitudes.  Yellow dashed lines show the magnetic domain walls in the stripe order phase.  (c) The magnitude of local moments in $\mu_B$ on the Cu  sites $i$ along $a$-axis defined as $|m_i|\equiv |n_{i\uparrow}-n_{i\downarrow}|\mu_B$, where $n_{i\sigma}$ is the occupancy of the $d_{x^2-y^2}$ Wannier orbital at site $i$ with spin $\sigma$.  (d) The $d_{x^2-y^2}$ occupancy on Cu atomic sites defined as $|n_i|\equiv |n_{i\uparrow}+n_{i\downarrow}|$.  Red dashed lines and blue solid lines represent the GAFM and stripe-order states, respectively.   Circles and diamonds show the results of the upper and lower layers.  
}
\label{fig:stripe_order}
\end{figure}

We begin with short-period magnetic orderings of the Cu magnetic moments, such as the non-magnetic, ferromagnetic, G-AFM, and A-AFM states.  Possible longer-period magnetic orders with spatial inhomogeneity, such as stripe order states, are considered further below.  Not surprisingly, the most energetically favorable magnetic order among the short-period orderings is the G-AFM order with antiparallel nearest-neighbor spins on Cu atoms as illustrated in Fig. \ref{fig:stripe_order}(a).  Other short-period meta-stable magnetic orders exhibit aligned nearest-neighbor spins, either in an intralayer ($\sim 40$ meV/Cu higher in energy) or interlayer ($\sim 2 $meV/Cu higher in energy) fashion, as discussed in the supplementary material \cite{supp}.  
Since the energy cost associated with changing the inter-bilayer spin alignment from antiparallel to parallel is negligible \cite{supp}, we will be concentrating below on the magnetic structure within a single bilayer. 

As we delve into the complex low-energy longer-period spin and charge orders, it will be crucial to be able to quantify the local Cu moment and electron count in a simple but precise manner.  Given the complexity of the large BSCCO supercell, we concentrate on the charge and spin of the electrons near the Fermi energy, i.e., the low-energy electronic behavior, so a low-energy Hamiltonian description will simplify the analysis.  As is well known for cuprates, the electronic bands at or near the Fermi energy are made from anti-bonding $\sigma$-type combinations of the Cu $3d_{x^2-y^2}$ and O $2p_{x/y}$ orbitals.  We find that other bands dominated by Bi, O $2p_z$, or other Cu $d$ orbitals are all at least 0.6 eV above or below the Fermi level (this is consistent with experimental knowledge that Bi-derived or other Cu $d$-derived bands do not appear in ARPES at the Fermi surface).  The minimal basis to describe the key bands is a ``one-band model'' \cite{feiner1996effective} where a single Wannier function is needed per Cu site: it has mixed anti-bonding Cu$d_{x^2-y^2}$/O$p_{x/y}$ character, and we construct it via the established maximal localization approach.  This means there is one Wannier function per CuO$_2$, and Appendix \ref{sec:threeband} shows a visualization of this Wannier function.  This one-band model is widely used to study various aspects of high-T$_c$ cuprates, e.g., spin and charge density waves, the pseudogap phase, and strange metal behaviors   \cite{leblanc2015solutions, huang2019strange, zheng2017stripe, fauque2006magnetic}.  However, instead of an idealized one-band model, Wannierization of the actual low-energy structure and electronic bands will automatically include symmetry-breaking effects into the one-band model (e.g., variation of on-site energies due to local modifications of bonding).  Not surprisingly, this Wannierization reproduced the DFT band structure about the Fermi energy to high accuracy (see the supplemental material \cite{supp}).  It is worth noting that another well-known effective model for the cuprates is the ``three-band model'' where one has Wannier functions for the Cu $d_{x^2-y^2}$ and both O$p_{x/y}$ so that one can describe the bonding explicitly as well as the low-energy O $2p$ $\sigma$-bonding-bands that are farther away from the Fermi energy.  Appendix \ref{sec:threeband} describes our three-band model and shows that it can also describe the low-energy bands accurately.  Given our focus on the band crossing the Fermi energy, we prefer the more economical one-band description.  

The one-band model has the advantage that the resulting electron counts are easy to understand: e.g., an undoped $d^9$ configuration for each Cu will correspond to one electron occupancy at each site; 25\% hole doping will correspond exactly to an average of 0.75 electrons per site (assuming only the bands described by the single-band model are hole-doped which is easily verified by direct comparison of the Wannierized band structure to the underlying DFT bands).  In essence, we have a localized basis that replicates the {\it ab initio} band structure near the Fermi energy. We employ this tight-binding representation to compute the band structure, band occupancies, and local occupancies of the Wannier orbitals. For a thermal Fermi-Dirac smearing of $0.01$eV ($\sim100$K), Figure \ref{fig:stripe_order}(c) presents the local moment magnitudes for the G-AFM state.  The moments are around 0.47$\mu_B$ with small modulations of about $\pm0.04\mu_B$ due to the superlattice modulation.  A similar modulation of $\pm0.03 e$ also manifests in the $d_{x^2-y^2}$ electron occupancies shown in Fig. \ref{fig:stripe_order}(d).  

We have discovered numerous longer-ranged stripe-ordered states that exhibit nearly degenerate energies with the G-AFM state.  While DFT-based stripe-ordered states were previously reported in LSCO \cite{anisimov2004computation, pesant2011dft} and YBCO \cite{zhang2020competing}, they have not been observed in BSCCO up to now.  For Bi-2212, we present a typical low-energy bond-centered stripe-ordered state in Fig. \ref{fig:stripe_order}(b), where the nearest neighbor spins crossing the dashed domain walls align in parallel, in contrast to the antiparallel alignment in the G-AFM state.  Remarkably, the total energy of this stripe-order state is only 1.9 meV/Cu higher than the G-AFM state.  In Appendix \ref{app:supp_stripe}, we tabulate eight distinct stripe order patterns with energies with 3 meV/Cu of the G-AFM state.  These stripe order states differ by their types and alignments of the domain walls, with only little energy differences.  The existence of all these energetically competing orders suggests the presence of strong spin fluctuations in the normal state, which can play an important role in superconducting pairing \cite{moriya2006developments, dahm2009strength, wakimoto2004direct, scalapino2012common}.  
Figs. \ref{fig:stripe_order}(c,d) show the local moment and $d_{x^2-y^2}$ occupancy of this stripe-order state.  The sites closer to the domain walls have lower occupancies and smaller local moments than the other sites.  These striped spin and charge orders present a modulation of $\pm0.2\mu_B$ for local moments and $\pm0.15$ for occupancies, significantly larger than the modulation caused by structural supermodulation captured in the G-AFM state.  Consequently, the formation of the stripe order has an electronic origin, but the precise location of the domain boundaries can be influenced by the superlattice modulation effect.  

Notice that the modulations of local moments and the electron occupancies are the same numerically.  This is because we have one Wannier orbital per site, and within band theory, forming a local moment requires an exchange splitting resulting in an occupied low-energy spin-majority orbital and empty high-energy minority-spin orbital: the doped holes go into the spin-majority orbital and reduce both local occupancy and magnetic moment simultaneously.  Hence,  the redistribution of doped holes strictly follows the change of spin structures in an intuitive manner.  

In addition to the Wannier basis described above, we have also analyzed the electronic structure of the stripe orders with the standard atomic projections output by the Vienna ab initio simulation package (VASP) software \cite{kresse1996efficiency, kresse1996efficient} in the supplementary material \cite{supp}.  The results are difficult to interpret due to the non-orthogonality of the standard VASP projections as explained in the supplement: the Cu magnetic moments show spatial modulations similar to those in Fig.~\ref{fig:stripe_order} but the total occupancies hardly vary from site to site;  additionally, there is a large oxygen contribution to the bands crossing the Fermi level.  In short, in contrast to our $d_{x^2-y^2}$ Wannier basis and the intuitive picture it provides, the standard VASP projections are unable to explain the stripe order easily and we do not discuss them further.  (The magnetic moments and electron counts for a three-band model are found in Appendix \ref{sec:threeband} which show qualitative agreement with our one-band model.)

\begin{figure*}[t]
\begin{center}
\includegraphics[scale=0.5]{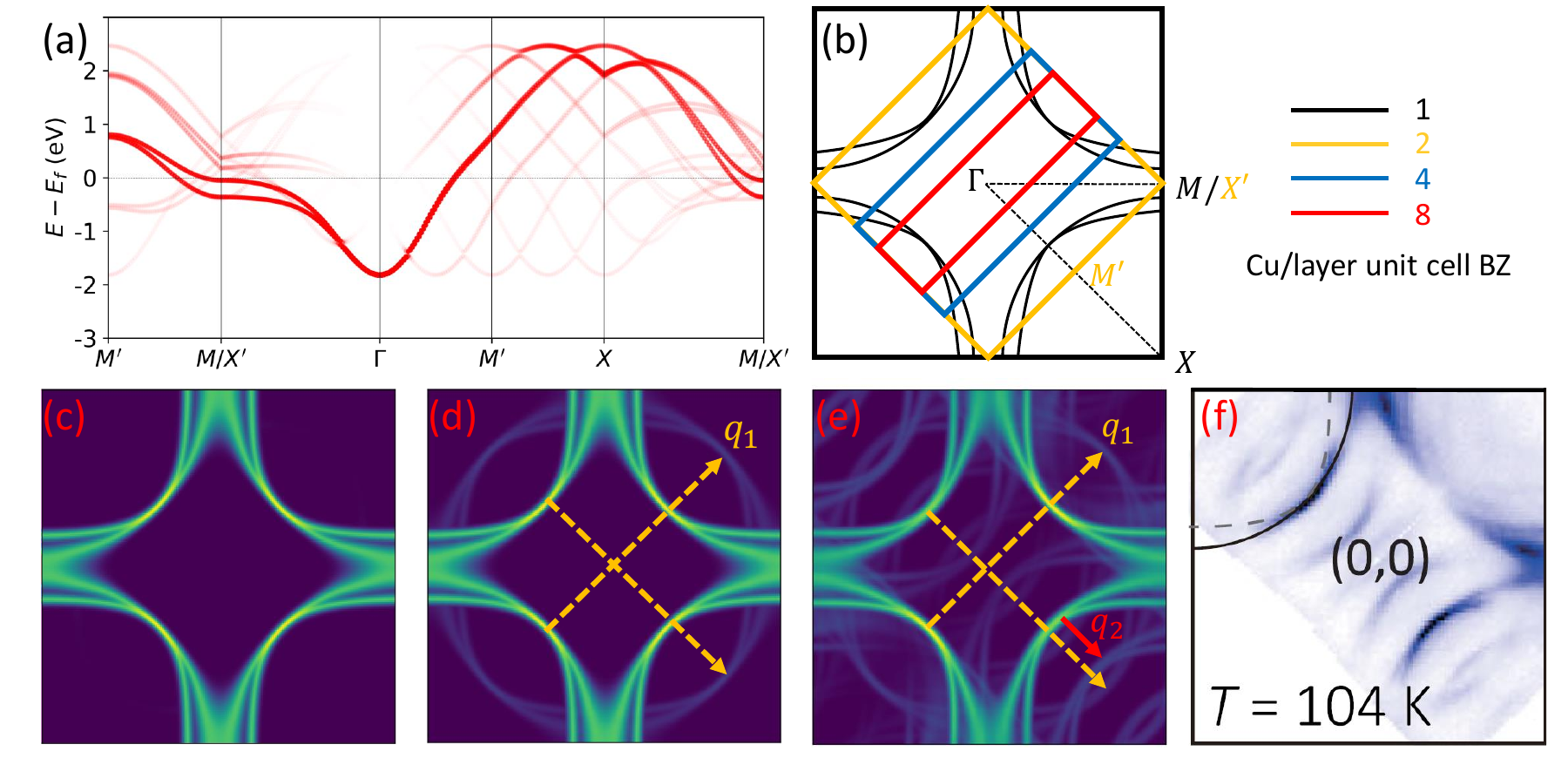}
\end{center}
\caption{
Electronic structures of the 25\% hole-doped Bi-2212 system with a non-magnetic state.  (a) The unfolded band structure of the 244-atom unit cell.  The opacity represents the spectral weight.  (b) Schematic Fermi surfaces are black curves.  The black, yellow, blue, and red rectangles represent the first BZ of the 1, 2, 4, and 8 Cu per layer unit cell, respectively.  (c-e) Unfolded Fermi surfaces for the symmetrized crystal, the orthorhombic distorted crystal of Fig. \ref{fig:low_symmetry}(b), and the hole-doped crystal of Fig. \ref{fig:244_crystal}, respectively.  The Fermi level in (c) and (d) is shifted by 25\% ``virtual hole doping'' to allow a fair comparison to the hole-doped Fermi surface in (e).  Yellow dashed ($q_1$) and red solid ($q_2$) arrows show two different coupling wave vectors.  (f) Measured ARPES Fermi surface at  $T=104$ K, adapted from experiments with similar hole doping level $\sim23\%$ \cite{he2021superconducting}.  Solid and dashed black curves highlight two distinct kinds of Fermi surface curves. 
}
\label{fig:244_bands}
\end{figure*}

\subsection{ARPES spectra and shadow bands}

The topology of the Fermi surface and the associated low-energy electronic spectrum, usually measured by ARPES \cite{damascelli2003angle}, provides important insights into the electronic properties of solids.  In particular, many materials exhibit a so-called ``shadow band'' (SB) Fermi surface, resembling a weak-intensity copy of their main band (MB) Fermi surface with certain shifted vectors in momentum space.  Depending on the system, these SB Fermi surfaces can, in principle, originate from any type of symmetry breaking such as electronic \cite{brouet2008angle}, magnetic \cite{rotenberg2005electron}, or structural \cite{mans2006experimental} origins.  The physical origin of the SB Fermi surface is crucial for understanding the physical properties of the material, but it is difficult to distinguish between the different possible origins from ARPES measurements alone.  

The Fermi surface of Bi-2212, as revealed by intensive ARPES studies, includes weaker intensity SB in addition to the main bands (see Fig. \ref{fig:244_bands}(f), adapted from an experiment \cite{he2021superconducting}).  These SBs can be described by two types of symmetry-breaking vectors.  One of them always aligns with the superstructural modulation direction, while the other is along $\pm(\pi,\pm\pi)$, coincident with the AFM ordering vectors.  This has led to a continued debate regarding the magnetic \cite{kampf1990spectral, langer1995theory, chubukov1995quasiparticle} versus the structural origin \cite{koitzsch2004origin, mans2006experimental, nakayama2006shadow, yu2020relevance} of the $\pm(\pi,\pm\pi)$ folding vector.  Direct theoretical interpretation of these SB Fermi surfaces using DFT has been lacking. 

Below, we predict the ARPES Fermi surface using DFT and many-body calculations to find the underlying physical mechanisms behind the emergence of SB.  We will demonstrate that the emergence of these two distinct types of SB is attributed to two distinct {\it structural} symmetry-breaking mechanisms while simultaneously clarifying how one should compute the ARPES spectrum in a theoretically consistent manner when building on  DFT output.

\subsubsection{Non-magnetic calculations}
\label{sec:NMcalcs}

Due to the strong spin fluctuations suggested by the numerous competing stripe orders and the G-AFM states, the normal state of the hole-doped system should not be described by a single magnetically-ordered configuration (i.e., a single Slater determinant).  In principle, an account of quantum spin fluctuations is needed for a comprehensive theoretical description including the magnetic susceptibility \cite{stepanov2018quantum} and spin correlations \cite{parsons2016site} in cuprates.  In practice, however, there is an established recipe whereby spectra computed using band theory for a non-magnetic electronic state of the cuprates are nicely comparable to ARPES Fermi surface \cite{damascelli2003angle, sobota2021angle}. 

In this subsection, we use this pragmatic approach and consider the electronic band structure of the non-magnetic state to see what can be learned.  The effect of spatial and temporal spin correlations and fluctuations on the computed ARPES spectra will be discussed in the following subsections.  As an added benefit, removing the spin degree of freedom allows us to focus exclusively on the effects of symmetry breaking by structural perturbations.  As demonstrated below, the non-magnetic state successfully accounts for many normal state spectroscopic properties, including the SB.  To facilitate comparison to experimental ARPES spectra of the Fermi surface, we employ a standard ``band-unfolding'' method \cite{ku2010unfolding,brouet2012impact} to project the band structure of a large supercell onto the primitive unit cell Brillouin zone.  This approach is known to reproduce spectral intensities observed in various ARPES experiments on many different materials qualitatively \cite{lin2011one, medeiros2014effects, tomic2014unfolding, zhu2018quasiparticle}.  The band structures and unfoldings discussed below are computed using the one-band Wannier basis described above in Sec. \ref{sec:244_magnetic_charge}.

The panels of Fig. \ref{fig:244_bands} display the unfolded band structures.  Fig. \ref{fig:244_bands}(a) shows the case of the 244-atom 25\% hole-doped system with a non-magnetic phase.  Around the antinodal region near M/X' in Fig. \ref{fig:244_bands}(a), we find two branches of flat bands below the Fermi level, which are split by the interlayer coupling in the bilayers, consistent with  ARPES experiments \cite{he2021superconducting, he2018rapid}.  

Fig. \ref{fig:244_bands}(e) exhibits the unfolded Fermi surface of this 25\% hole-doped system.  The curves with the highest intensity contain two easily visible curves corresponding to the solid and dashed black curves of Fig. \ref{fig:244_bands}(f) from the ARPES measurement.  Aside from these main curves, the ARPES Fermi surface exhibits a complicated set of shadow bands with lower intensities which are also reproduced in Fig. \ref{fig:244_bands}(e). Given the highly satisfactory theoretical results, we will conduct an analysis to demonstrate that the different sets of shadow bands have two distinct physical origins.

We begin with a symmetrized model of the Wannier tight-binding model of undoped BSCCO with orthorhombic distortions: the spatially inhomogeneous onsite energies and hoppings introduced by structural distortion are symmetrized to their respective mean values.  
As a result, the non-magnetic state of this model can be described in a small primitive cell with one Cu atom in each CuO$_2$ layer, due to the full translational symmetry of this symmetrized model.  Consequently, the unfolded Fermi surface of this model, shown in Fig. \ref{fig:244_bands}(c), only contains the main Fermi curves without any shadow bands.  (The Fermi levels of the undoped crystals are lowered manually to achieve 25\% hole doping.)  Starting from this clean Fermi surface, we successively add complexity to the crystal to see the emergence of the shadow bands.

Next, we turn to the effect of orthorhombic distortions on the Wannier tight-binding model given by the crystal in Fig. \ref{fig:low_symmetry}(b) without symmetrization.  Similarly, 25\% hole doping is introduced by lowering the Fermi level manually.  This model exhibits an approximately circular-shape shadow band as shown in Fig. \ref{fig:244_bands}(d).  This circular shape shadow band arises from the symmetry breaking of in-plane hoppings due to the orthorhombic distortions, as depicted in Fig. \ref{fig:low_symmetry}(b).  Previous experiments have observed these shadow bands and suggested their likely structural origin \cite{koitzsch2004origin, mans2006experimental, nakayama2006shadow} without providing a specific microscopic picture.  We observe that the distortions in the Bi-O layers enlarge the unit cell and introduce an inter-band coupling at wave vectors $q_1 = \pm(\pi,\pm\pi)$.  This coupling leads to folding from the main bright curves to the shadow bands as indicated by the yellow dashed arrows in Fig. \ref{fig:244_bands}(d) and Fig. \ref{fig:244_bands}(e).  

Finally, we study the Wannier model for the large supercell shown in Fig. \ref{fig:244_crystal} that includes both the Bi-O layer distortions and the superlattice modulation.  Consequently, on top of the circle-like shadow band, there is another type of shadow band involving the wave vector $q_2 = \pm(\pi/4,-\pi/4)$ shown in Fig. \ref{fig:244_bands}(e).  This type of shadow band, often referred to as a ``superstructure'' in previous studies \cite{liu2019evolution}, has long been attributed to a post-emission modulation effect from Bi-O layer buckling \cite{yu2020relevance}.  Prior experiments have also shown that this type of shadow band fades away when the crystal modulation is gradually reduced by Pb doping \cite{liu2019evolution}.  Our calculation presents a consistent and simple microscopic picture: the oxygen dopants and associated superlattice modulations further enlarge the unit cell and introduce an additional coupling at $q_2$.  This coupling creates a further folding (illustrated by the red arrow) from the main bright curves to another set of shadow bands.  Note that, in both experiments and our theory, this coupling only occurs in one crystalline direction due to the superlattice modulations being solely along the $a$-axis as shown in Fig. \ref{fig:244_crystal}.   

Concluding this subsection, results based on the non-magnetic electronic state clearly show that lattice distortions alone can give rise to the shadow bands observed on the Fermi surface of BSCCO and that the matching of theory and experiment is of very high quality.  Many-body effects and complex electronic fluctuations are not necessary to describe the shadow bands.  However, since many-body effects and fluctuations exist in the actual material, describing the material using a non-magnetic state electronic state is highly inconsistent from a theoretical viewpoint.  As a simple example of the inconsistency, the non-magnetic state is much higher in energy than any of the magnetically ordered phases in our (and prior) DFT calculations: why is an ARPES spectrum computed using an unphysical high-energy state so accurate?

\subsubsection{Band theory treatment: static inhomogeneous local moments}

DFT calculations (ours, as well as prior work, \cite{zhang2020competing}) show that there are many competing low-energy states involving charge and spin ordering.  Thus the simplest step going beyond the non-magnetic calculation is to stay with the DFT framework (i.e., band theory) and assume that the actual material has static local moments on each atomic site and is spatially inhomogeneous microscopically: different parts of the material have different local charge/spin orderings.  This state is often known as the spin glass state.  If experimental measurements average spatially over these inhomogeneities, one can attempt to describe the fact that.  Like other cuprate superconductors, $x=0.25$ hole-doped BSCCO shows strong spin fluctuations \cite{lyons1988dynamics, brom2003magnetic, zhu2023spin, lyons1988spin} and is paramagnetic (PM) \cite{sachdev2003colloquium, yoshizaki1988superconducting, kaiser2012curie} in experiments.  Hence, the static spatial fluctuations of the charges/spins from the band theory should average to give a globally uniform and paramagnetic structure.  The key point is that because band theory is based on a single Slater-determinant which precludes dynamic (temporal) electronic fluctuations, the fluctuations in the actual material must necessarily be described, within band theory, as static ones that are distributed inhomogeneously in space.  In this section, we will see that this approach leads to poor or problematic predictions of ARPES spectra and that a method going beyond band theory will be needed for a consistent and accurate description.

\begin{figure}[t]
\begin{center}
\includegraphics[scale=0.53]{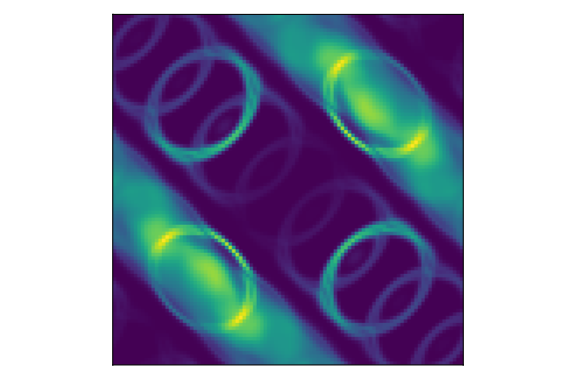}
\end{center}
\caption{Thermally averaged unfolded Fermi surface of eight representative low-energy stripe-order states and the G-AFM state from the DFT calculations.  Labels of high-symmetry $k$-points are the same as Fig. \ref{fig:244_bands}. }
\label{fig:stripe_FS}
\end{figure}

The simplest approach to dealing with inhomogeneity is to take a large library of DFT-computed low-energy structures and compute a thermally-averaged ARPES spectrum: for each spin/charge-ordered state, one computes the ARPES spectrum using band-unfolding and has it contribute to the average spectrum with a weight given by its thermal Boltzmann probability based on the energy of that state per unit cell.  The physical picture is that each unit cell of the material is effectively independent of its neighboring cells so that the spatial inhomogeneity can be replaced by a thermal average over a single unit cell's multiple possible states. In other words, one assumes that the domain size of each phase is on the order of a unit cell \cite{markiewicz2019first}.  We have performed this calculation using 8 different low-energy stripe order states as well as the G-AFM state for the 244-atom unit cell using a thermal energy of $k_BT=0.01$ eV.  The resulting ARPES spectrum is shown in Fig. \ref{fig:stripe_FS} and has little relation to the experimental observations. 

Given this problematic spectral prediction, the above approach can be improved by relaxing the assumption of the fixed domain sizes: one should make a more faithful model that automatically chooses domain sizes and distributions based on total energy minimization in a large supercell, and then compute the associated ARPES spectrum of this large supercell that contains representative inhomogeneous charge/spin distributions \cite{wang2020understanding}.  It turns out (below) that one needs beyond 2,000 Cu sites in the supercell to converge the resulting ARPES spectra versus supercell size.  Hence, this approach is not presently feasible using DFT due to its computational expense, and a surrogate model is needed.  In the following, we will use a simple tight-binding approach based on Wannierization of the DFT band structure to arrive at a computationally tractable surrogate model.

Specifically, we consider a single-band Hubbard model treated within Hartree-Fock (HF) theory for a single 2D CuO$_2$ plane in BSCCO (the same Wannier basis from Section \ref{sec:244_magnetic_charge}), so we have one $d_{x^2-y^2}$ Wannier function for each Cu site.   The total HF energy is 
\begin{multline}
E_{tot}^{HF} = -\sum_{ij\sigma} t_{ij}\langle \hat c^\dag _{i\sigma}\hat c_{j\sigma}\rangle \\
+ \sum_{i\sigma} \epsilon_i \langle\hat n_{i\sigma}\rangle + U\sum_i \langle \hat n_{i\uparrow}\rangle \langle \hat n_{i\downarrow}\rangle
\label{eq:EtotHF}
\end{multline}
where $\hat c_{i\sigma}$ is the electron (fermion) annihilation operator for site $i$ with spin $\sigma$, the number counting operator is $\hat n_{i\sigma}=\hat c^\dag_{i\sigma}\hat c_{i\sigma}$, $t_{ij}$ is the hopping parameter between sites $i$ and $j$, and $\epsilon_i$ are on-site energies.  The operator expectations are over a single Slater determinant wave function found by minimization of the total energy or equivalently self-consistent solution of the HF single-particle Hamiltonian, which for spin channel $\sigma$ is 
\begin{equation}
\hat H^{HF}_\sigma = -\sum_{ij} t_{ij} \hat c^\dag _{i\sigma}\hat c_{j\sigma} +\sum_{i} \epsilon_i \hat n_{i\sigma}+ U\sum_i \langle \hat n_{i-\sigma}\rangle \hat n_{i\sigma}
\label{eq:HHF}
\end{equation}
where $-\sigma$ denotes the opposite spin from $\sigma$.

For BSCCO, we find that a high precision reproduction of the DFT band structure requires only three key types of hopping parameters which are $t=t_{\avg{100}},t'=t_{\avg{110}}$ and $t''=t_{\avg{200}}$ where $\avg{100}$ is a nearest-neighbor hopping (along the Cu-O-Cu direction), etc.  The Wannierization of the full 244-atom unit cell including the crystal modulation distortion yields us the hopping parameters $t_{ij}$ as well as the site-dependent $\epsilon_i$; the hopping parameters, when averaged over all appropriate pairs in the unit cell, have the values $t=-0.473$, $t'=0.088$, and $t''=-0.091$ eV.  (To simulate a high-symmetry structure, we symmetrize our model by setting all $\epsilon_i=0$ and all hoppings to these averaged values.)  These hopping parameters are qualitatively consistent with the effective Hamiltonians in prior DFT studies on high-symmetry unrelaxed crystals \cite{moree2022ab, schmid2023superconductivity, ohgoe2020ab, hirayama2019effective}.  In addition, due to the structural distortions and superlattice modulation, these hoppings are modulated spatially by about $\pm 0.015$eV, and the onsite energies are modulated by about $\pm 0.032$eV.   The interaction parameter $U$ is chosen so that the HF total energy difference between the non-magnetic and GAFM phases at $x=0.25$ reproduces the total DFT energy difference between those phases.  The resulting value is $U=3.1$ eV or $U/t\approx 7$, which is consistent with the effective Hamiltonian from prior works \cite{moree2022ab, schmid2023superconductivity, ohgoe2020ab, nilsson2019dynamically, hirayama2019effective}, and is very reasonable compared to prior studies of cuprate systems using one-band Hubbard models \cite{sheshadri2023connecting, dalla2012unified, zheng2017stripe}.

\begin{table}
\begin{tabular}{c|rcc|rcc}
  & \multicolumn{6}{c}{Method}\\
  & \multicolumn{3}{c}{HF} & \multicolumn{3}{c}{SSSB}\\
 Phase &\ \ $E$ \ \ & \ \ \ $|m|$ \ \ \ & \ \ \ $n$ \ \ \ & \ \ $E$ \ \ & \ \ \ $|m|$ \ \ \ & \ \ \ $n$ \ \ \ \\
\hline
$x=0$ & & & & &\\
NM & 0 & 0 & 1 & 0 & 0 & 1\\
FM & -27 & 0.8 & 1 & -0.2 & 0.04 & 1\\
AFM & -236 & 0.9 & 1 & -56 & 0.8 & 1\\
& & & & & & \\
$x=0.25$ & & & & &  \\
NM & 0 & 0 & 0.75 & 0 & 0 & 0.75\\
FM & -74 & 0.6 & 0.75 & -0.2 & 0.06& 0.75\\
AFM & -95 & 0.5 & 0.75 & -3 & 0.3 & 0.75\\
stripe & -106 & 0.4-0.7 & 0.66-0.84 & -2 & 0.1-0.2 & 0.74-0.75\\
random & -117 & 0.0-0.8 & 0.49-0.94 & -3 & 0.0-0.4 & 0.74-0.77\\
\end{tabular}
\caption{Key ground-state properties for the undoped $x=0$ and hole-doped $x=0.25$ 2D single-band Hubbard model based on Hartree-Fock (HF) and the single-site slave-boson (SSSB) methods with $U=3.1$ eV and using hopping parameters extracted from Wannierization of DFT band structures.  Results are based on calculations on a $32\times32$ lattice with periodic boundary conditions and thermal energy of $k_BT=0.01$ eV.  The phases describe the static spatial distributions of charge and spin: spatially uniform non-magnetic (NM), spatially uniform ferromagnetic (FM), checkerboard (G-type) anti-ferromagnetic (AFM), period-4 stripe phase of alternating FM and AFM lines (stripe), and a representative low-energy state started from a random seed (random).  The columns are the total energy per electron $E$ in meV, the magnitude of the local magnetic moment $|m|$ (where $m_i=n_{i\uparrow}-n_{i\downarrow}$ for each site $i$), and the local electron count $n$ ($n_i=n_{i\uparrow}+n_{i\downarrow}$). For spatially non-uniform phases, the minimum and maximum of $|m_i|$ and $n_i$ are provided as a range.  For each doping level, the NM phase is chosen as the reference zero energy.}
\label{tab:HFSBenergies}
\end{table}

\begin{figure*}[t!]
\begin{center}
\includegraphics[scale=0.3]{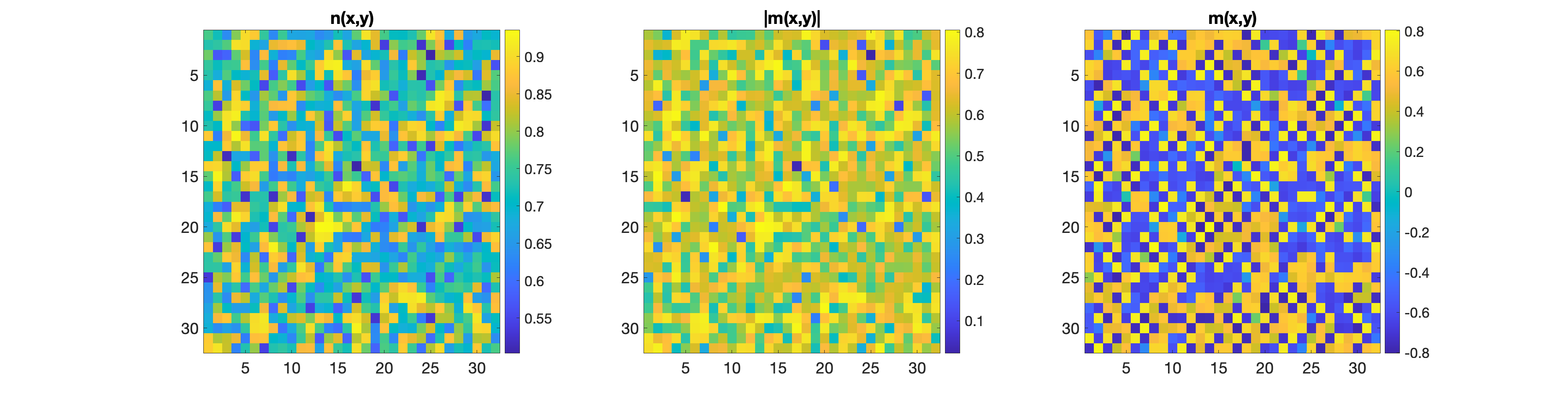}
\end{center}
\caption{Self-consistent local electron density and magnetic moments from a HF treatment of the 2D one-band Hubbard model for a $32\times32$ periodic 2D lattice with $U=3.1$ eV, a thermal smearing of $k_B T=0.01$ eV, and hole-doping $x=0.25$.  This particular solution was started from a random initial guess and has a lower HF total energy than any ordered phase (see Table \ref{tab:HFSBenergies}).  Left: local electron density  $n=n_{i\uparrow}+n_{i\downarrow}$ for each site $i=(x,y)$.  Middle and right: magnitude and signed value of the local magnetic moment $m_i=n_{i\uparrow}-n_{i\downarrow}$.  Note the complex mixture of AFM and FM arrangements domains and how sites with large magnetic moments $|m_i|$ also have high electron density $n_i$.  }
\label{fig:HF2ddensityplots}
\end{figure*}

We solve for HF solutions using a simple self-consistent field (SCF) method: we diagonalize the HF one-particle Hamiltonian of Eq.~(\ref{eq:HHF}) using k-point sampling of a large periodic supercell, compute the resulting electron densities $\langle \hat n_{i\sigma}\rangle$ using small thermal smearing of $0.01$ eV (for numerical stability of the SCF calculation), and iterate to convergence.  We begin with a variety of different initial seed densities (e.g., ferromagnetic, AFM, stripes, or random) to arrive at solutions of different symmetries.  For each solution, we compute the ARPES spectrum by computing a dense sampling of wave vectors of the supercell, finding the states at the Fermi energy, and projecting their wave functions onto plane waves.  To converge the ARPES spectra with the above thermal smearing, our 2D lattices must be of minimum size $32\times32$ (i.e., 1,024 Cu sites in one CuO$_2$ layer).  In addition to sampling a large supercell, we thermally average the ARPES spectra over $\sim 200$ different low-energy HF solutions to simulate averaging over a large inhomogeneous materials system.

Table \ref{tab:HFSBenergies} (left side) shows total HF energies per electron and local magnetic moments for several low-energy solutions for undoped $x=0$ and hole-doped $x=0.25$ systems.  Not surprisingly, the undoped $x=0$ system has a checkerboard AFM ground state with a large stabilization energy. For the hole-doped case, the lowest energy states (represented by the ``random'' row) have a complex, inhomogeneous spatial distribution of spin and charge: we obtain a large number of nearly degenerate solutions that are qualitatively similar in that they show mixed domains of AFM and FM with meandering boundaries.  Figure \ref{fig:HF2ddensityplots} shows an example of the magnetic moment and electron count distribution of a low-energy HF solution (the ``random'' entry of Table \ref{tab:HFSBenergies}).  We note a few facts from the figure:  the lattice has a mixture of FM and AFM domains, each domain is a few unit cells wide (typically 5 or less), and lattice sites with high electron count $n$ have a larger magnetic moment $|m|$ (and vice versa).  In other words, within HF, the doped holes choose to distribute themselves inhomogeneously in space to lower the total energy, and the sites to which they segregate have weak magnetic moments.

\begin{figure}[t]
\begin{center}
\includegraphics[scale=0.46]{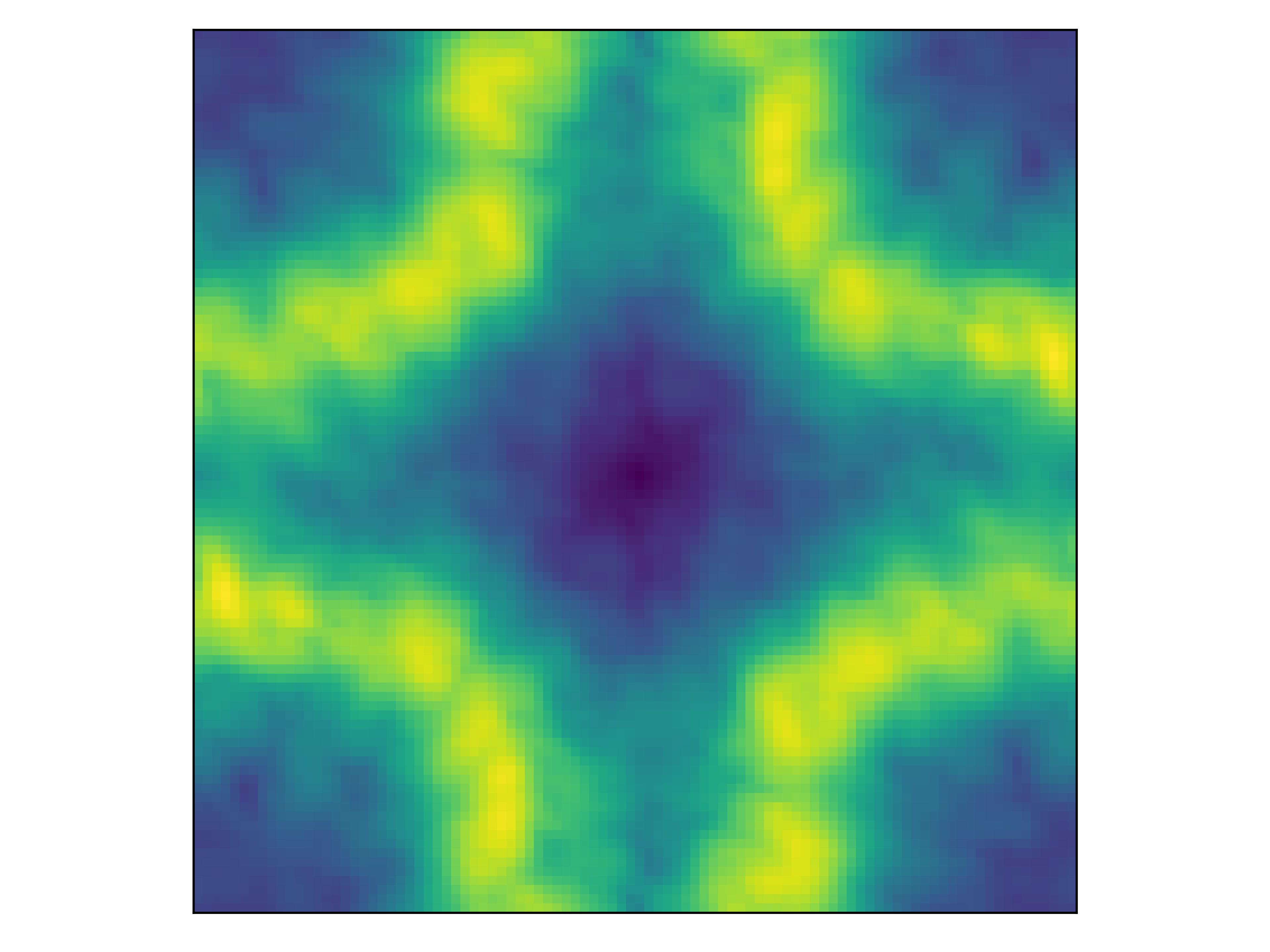}
\end{center}
\caption{Band-unfolded Fermi surface for the low-energy, spatially inhomogeneous HF solution corresponding to Fig.~\ref{fig:HF2ddensityplots}.  The rough outlines of an NM-like spectrum similar to the experimental and theoretical findings in Fig.~\ref{fig:244_bands} are visible, but the spectral weight around the Fermi level is too broad for a faithful comparison or for visualizing any shadow bands.}
\label{fig:HFbandfold}
\end{figure}

Unfortunately, when we compute the unfolded band structure at the Fermi level for such low-energy HF solution associated with Fig.~\ref{fig:HF2ddensityplots}, we find results shown in Fig.~\ref{fig:HFbandfold}.  Overall, the spectrum is qualitatively similar to the non-magnetic theoretical as well as experimental spectra from Fig.~\ref{fig:244_bands} as it has visible Fermi arcs. This makes sense because a large simulation cell with an electron distribution like that of  Fig.~\ref{fig:HF2ddensityplots} is non-magnetic upon spatial averaging, and the wave functions used to compute the spectrum in Fig.~\ref{fig:HFbandfold} are plane waves of constant amplitude that sample the entire simulation cell.  However, this HF-derived spectral weight is far too broad in wave vector (momentum) space to make a meaningful comparison to the experiment or to even visualize the shadow bands.  In retrospect, the reason is straightforward: the static and inhomogeneous electronic distribution shown in Fig.~\ref{fig:HF2ddensityplots} contains domains whose size is a few primitive unit cells, so we would expect upon Fourier transformation to find broad structures in wave vector space.  

Thus, increasing the size of the simulation cell when averaging spectra derived from static spatial inhomogeneity does improve the comparison to the experiment (compare Fig.~\ref{fig:stripe_FS} to Fig.~\ref{fig:HF2ddensityplots}), but there is still quite a distance to go before claiming to understand the electronic structure that underlies the experimental ARPES spectra.  Since the band theory approach assumes an electronic state that is described by a static spatial pattern, we have to abandon this underlying assumption to make progress.

\subsubsection{Many-body treatment: dynamic fluctuations}
\label{sec:clustersb}
To go beyond band theory, we need a theoretical framework that allows us to include the basic physics of dynamic electronic fluctuations in a computationally efficient manner.  As noted above, the fact that the 25\% hole-doped BSCCO is paramagnetic (PM) \cite{sachdev2003colloquium, yoshizaki1988superconducting, kaiser2012curie} argues that each local spin moment fluctuates dynamically (i.e., in time) \cite{zaanen1996dynamical, lyons1988dynamics, lyons1988spin, brom2003magnetic, zhu2023spin}.  Prior research has also suggested a close relation between spin fluctuations and high-temperature superconductivity \cite{moriya2000spin, zhu2023spin}.  In this section, we perform a many-body study of the system and arrive at two findings: (a) a many-body treatment removes the energetic drive towards a static, inhomogeneous electronic distribution and instead produces a spatially almost uniform static (i.e., time-averaged) spin/charge distribution where the residual spatial inhomogeneity is only introduced by structural distortions; and (b) while many-body effects narrow the electronic bandwidths when compared to the NM DFT calculations, they have little effect on the Fermi surface including the shadow bands. 

There are a number of different many-body solid-state methods when one wishes to go beyond mean-field theory.  Considering the sizes of the simulation cells we are dealing with even for the simple 2D single-band Hubbard model, we will use computationally efficient slave-boson methods that we have developed recently as both single-site methods~\cite{georgescu2015generalized,georgescu2017symmetry,georgescu_boson_2021} and an accurate cluster-based method~\cite{jin2023bond}.  In these approaches, the electron annihilation operator $\hat c_{i\sigma}$ is replaced by the product $\hat f_{i\sigma}\hat O_{i\sigma}$ where we have a chargeless fermion represented by the fermionic annihilation operator (called a spinon) $\hat f_{i\sigma}$ and a charged auxiliary or slave boson represented by its lowering operator $\hat O_{i\sigma}$.  The two subsystems are separated by approximating the ground-state density matrix $\hat \rho$ of the combined spinon+boson system by the product form $\hat\rho_f\times\hat\rho_s$ where each subsystem has its own density matrix ($\hat\rho_f$ for spinons and $\hat \rho_s$ for the slave bosons).  The ground state is found by minimizing the total energy 
\begin{multline*}
E_{tot}^{SB} = -\sum_{ij\sigma} t_{ij}\langle \hat f^\dag _{i\sigma}\hat f_{j\sigma}\rangle_f\langle \hat O_{i\sigma}^\dag\hat O_{j\sigma} \rangle_s \\
+ \sum_{i\sigma} \epsilon_i \langle\hat n_{i\sigma}\rangle_f + U\sum_i \langle \hat n_{i\uparrow}\hat n_{i\downarrow}\rangle_s
\end{multline*}
where the $f$ or $s$ subscripts mean an average using the associated density matrix $\hat\rho_f$ or $\hat\rho_s$.  We note that the interaction term is treated correctly within the slave-boson description (i.e., the product of densities is not factorized as in HF in Eq.~(\ref{eq:EtotHF})).  Total energy minimization for this approach means one must solve two coupled problems: a set of non-interacting spinons (fermions) moving on a lattice whose hoppings are modulated by the slave particles, and a set of interacting slave modes that are moving on a lattice.  In practice, the slave problem is approximated by solving small clusters: the slave Hamiltonian of a single site or a small cluster of interacting sites is solved exactly and is coupled to an effective bath self-consistently determined from the averaged properties of the site or cluster.  

To compute electronic spectra at the Fermi energy, one considers the spinon Hamiltonian given by
\begin{equation}
    \hat H_f = -\sum_{ij\sigma}t_{ij} \langle \hat O_{i\sigma}^\dag\hat O_{j\sigma} \rangle_s \hat f^\dag _{i\sigma}\hat f_{j\sigma} + \sum_{i\sigma} (\epsilon_i+B_{i\sigma}) \hat n_{i\sigma}
    \label{eq:Hspinon}
\end{equation}
where the auxiliary ``magnetic'' fields $B_{i\sigma}$ permit static symmetry-breaking solutions of the electronic ground state \cite{georgescu2017symmetry,jin2023bond}: part of the minimization process is to find the $B_{i\sigma}$ values that minimize the total energy.  The spinon Hamiltonian $\hat H_f$ describes a set of non-interacting electrons moving on a lattice whose tight-binding hoppings are renormalized by expectations values from the slave sector.  The spectral properties of the interacting spinon+slave system at the Fermi energy are found by computing them at the Fermi energy of the spinon problem (e.g., by diagonalization of $\hat H_f$) \cite{riegler2020slave}.  Note that this approach can reproduce the Hartree-Fock predictions if total energy minimization results in $\langle \hat O_{i\sigma}^\dag\hat O_{j\sigma} \rangle_s=1$ and $B_{i\sigma}=U\langle \hat n_{i-\sigma}\rangle$, but it has a wider set of solutions with $\langle \hat O_{i\sigma}^\dag\hat O_{j\sigma} \rangle_s\ne1$ with dynamical spin/charge fluctuations that may have lower energy.

We begin with the simplest slave-boson method where the interacting problems being solved involve each site separately.  We have performed a large number of calculations on $32\times32$ lattices for the above Hubbard model (with periodic boundary conditions), minimizing the total energy of each system when starting with various initial electronic seed states (NM, FM, AFM, stripe, and random).  Table~\ref{tab:HFSBenergies} provides total energies and basic observables for the undoped and hole-doped 2D Hubbard lattices.  Superficially, like HF, static electronic symmetry breaking for slave bosons also lowers the energy.  

However, the main message of the table is the vastly different energy scale for stabilization of static electronic symmetry breaking: the energy lowering in the slave-boson calculations is one to two {\it orders of magnitude} smaller than in HF, especially for the hole-doped case; the magnitude of the magnetic moments and amplitude of spatial variations in electron density are also much weaker in the slave-boson case.  A band theory method such as HF will lower total energy primarily by reducing the repulsive interaction energy via symmetry breaking (e.g., by forming static local magnetic moments with $\langle n_{i\uparrow}\rangle\ne\langle n_{i\downarrow}\rangle$); the strength of this tendency is controlled directly by the interaction parameter $U$.  In the single-site slave-boson approach, the many-body solution for each site explicitly describes the empty, singly, and doubly occupied configurations, so the interaction energy can be lowered by reducing the contribution of the doubly occupied states without any symmetry breaking.  Hence, the weak residual static symmetry breaking is driven by the system trying to take advantage of weaker inter-site interactions (e.g., superexchange).

\begin{figure}[t]
\begin{center}
\includegraphics[scale=0.52]{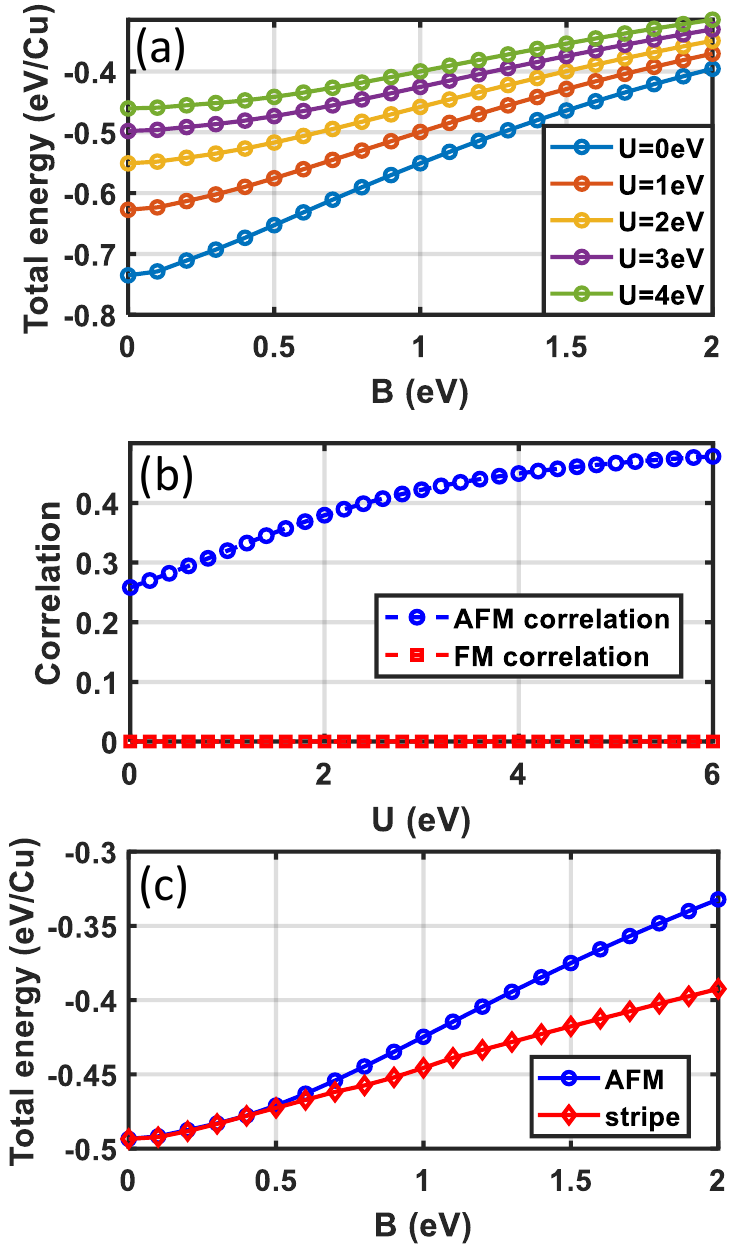}
\end{center}
\caption{
Cluster slave-boson results for hole-doped 2D Hubbard model derived from Wannierzation of the 244-atom unit cell. (a) The total energy as a function of the auxiliary field strength $B$ applied in 
 staggered static AFM pattern.  The local interaction $U$ of the Hubbard model is varied from 0 to 4 eV.  (b) Nearest neighbor spin correlations of the PM ground state for $U=3.1$ eV: AFM correlation is defined as $\avg{\hat{N}_{i\uparrow} \hat{N}_{j\downarrow}+\hat{N}_{i\downarrow} \hat{N}_{j\uparrow}}_s$ while FM correlation is defined as $\avg{\hat{N}_{i\uparrow} \hat{N}_{j\uparrow}+\hat{N}_{i\downarrow} \hat{N}_{j\downarrow}}_s$ (both are averaged over all nearest neighbor pair of sites $ij$).  (c) The total energies of static AFM and stripe phases as a function of the magnitude of the auxiliary $B$ field for $U=3.1$ eV.  The static AFM state is induced by applying a staggered $B$ for all sites. The stripe state is induced by applying $B$ to the strong local moment sites along their corresponding spin directions as shown in Fig. \ref{fig:stripe_order}(b) (marked by large black arrows); the auxiliary fields at the weak local moment sites are optimized to minimize the total energy at each data point.  
}
\label{fig:slaveboson}
\end{figure}

If one enlarges the size of the interacting cluster being treated, the driving force for static symmetry breaking will further weaken, and this is what we find.  When we enlarge our interacting problem to involve two neighboring sites and use our cluster slave-boson approach \cite{jin2023bond}, we find that energy minimization of the ground states leads to a solution with no static symmetry breaking. Here, we take the DFT-calculated 244-atom $x=0.25$ unit cell electronic structure and perform Wannierization to build the Hubbard model.  Therefore, each interacting slave cluster is solved in its own unique local supermodulation crystalline environment as dictated by the extract $\epsilon_i$ on-site energies and hoppings $t_{ij}$. Figs.~\ref{fig:slaveboson}(a) and (c) show that the total ground state energy increases when either a static AFM or stripe pattern of magnetic moments is imposed: the lowest energy solution shows no static symmetry-breaking ($B=0$).  However, as shown in Fig.~\ref{fig:slaveboson}(b), there are strong AFM correlations between neighboring sites as one expects from the physics of super-exchange.  In brief, energetically favored spin-spin correlations are fully dynamic at this doping level.  We refer to this solution as a paramagnetic state (PM).  

\begin{figure}[t]
\begin{center}
\includegraphics[scale=0.46]{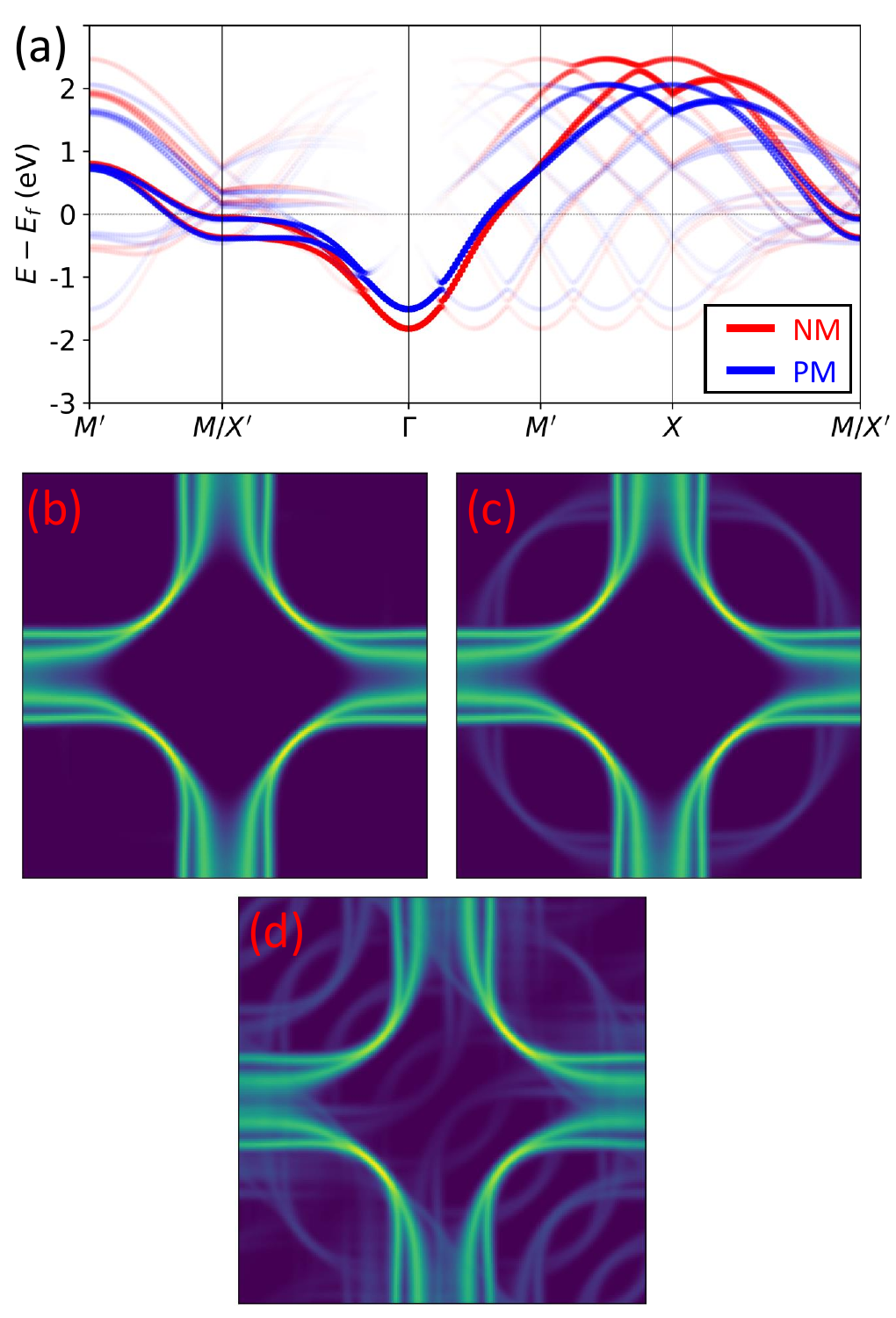}
\end{center}
\caption{
Electronic structures of 25\% hole-doped Bi-2212 system with PM ground state.  (a) The unfolded band structure of the 244-atom unit cell.  Red and blue curves represent DFT NM bands and spinon bands for the PM state, respectively.  (b-d) The PM state unfolded Fermi surface of the symmetrized crystal, the orthorhombic distorted crystal of Fig. \ref{fig:low_symmetry}(b), and the hole-doped crystal of Fig. \ref{fig:244_crystal}, respectively.  Similarly, the Fermi level in (b) and (c) is shifted by 25\% virtual hole doping to allow a fair comparison to the
hole-doped Fermi surface in (d).  These Fermi surfaces are very similar to the NM states in Fig. \ref{fig:244_bands}(c-e).  
}
\label{fig:PMstate}
\end{figure}

The spatial homogeneity of the minimum energy many-body solution means that when computing the Fermi surface, the appropriate spinon Hamiltonian $\hat H_f$ (Eq.~\ref{eq:Hspinon}) has $B_{i\sigma}=0$ and renormalization factors $\langle \hat O_{i\sigma}^\dag\hat O_{j\sigma} \rangle_s$ that are essentially constant throughout the unit cell (they vary by less than 10\% among the various slave clusters $ij$ due to the crystal distortions).  Therefore, the PM Fermi surface ends up looking very similar to the NM DFT Fermi surface from section \ref{sec:undoped}: Fig.~\ref{fig:PMstate}(a) compares the two band structures showing that the modest but quantitative renormalization (narrowing) of the spinon bands.  In addition, the interactions have little effect near the Fermi surface, and this is verified explicitly in Fig.~\ref{fig:PMstate}(b,c,d) which can be compared to the DFT NM results in Fig.~\ref{fig:244_bands}.  In particular, Fig.~\ref{fig:PMstate}(b) shows the Fermi surface of the symmetrized crystal in the PM state.  While the dynamical spin fluctuations and correlations are properly described in this study, the many-body effect does not directly cause any shadow bands.  However, the shadow bands appear in Fig.~\ref{fig:PMstate}(c) and (d) where additional crystal distortions are introduced.  Therefore, like the NM DFT results, the shadow bands originate from the crystal structure of the materials and do not originate in electronic or magnetic interactions.

In summary, the reason the DFT NM Fermi surfaces do so well at reproducing the experimental Fermi surfaces lies in the fact that the actual material system at $x=0.25$ has a PM ground state without static local moment \cite{sachdev2003colloquium, yoshizaki1988superconducting, scalapino2012common}: the DFT NM is the best band theory solution mimicking this fact despite being a high-energy solution within the DFT framework.  Of course, the electrons do have significant spin-spin correlations that lower the total energy, but in this case, the correlations are purely dynamic and are not reflected in static spatial symmetry breaking. Hence, while the band structure calculations systematically and sensibly predict low-energy states with spatially inhomogeneous static patterns of charge and spin, the actual system dynamically and quantum mechanically fluctuates among these ``snapshots'' during the experimental photoemission process leading to an effectively NM-looking final Fermi surface (and this is not the same as averaging over spectra from a large number of static snapshots).  Hence, we can now explain the success of the recipe of using NM DFT calculations to describe band structures of cuprate phases that do not display long-range spin/charge order in experiments.

\section{Modulation of correlation strength}

\begin{figure}[t]
\begin{center}
\includegraphics[scale=0.58]{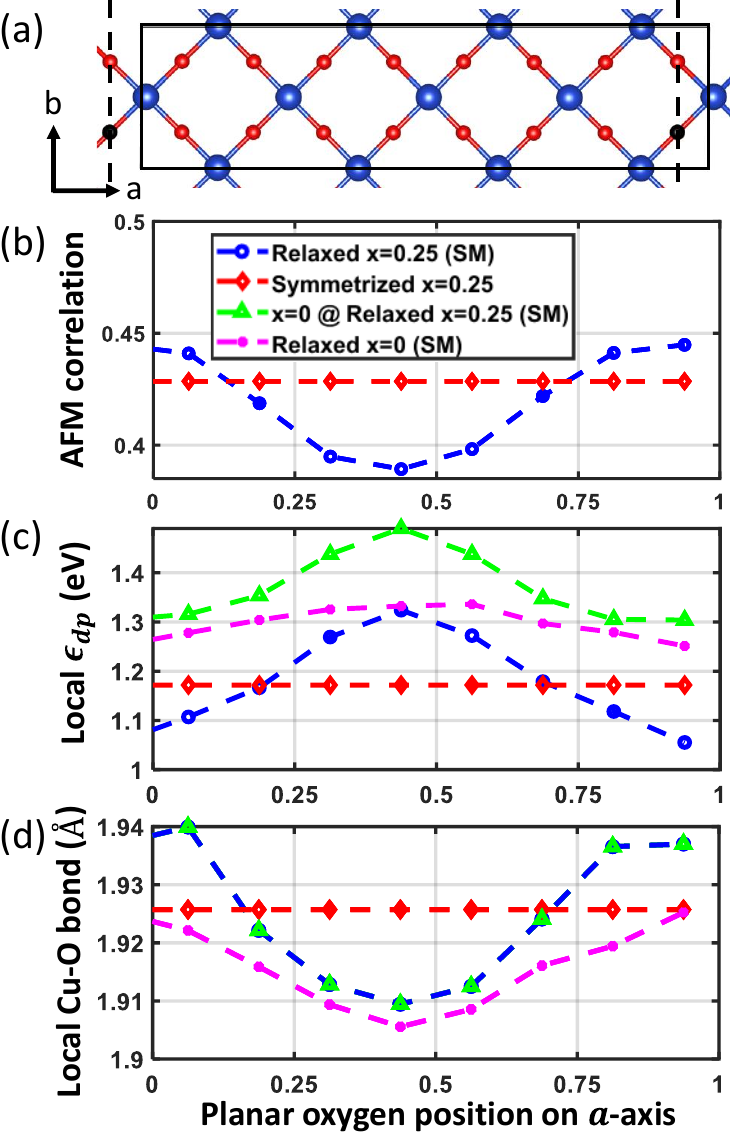}
\end{center}
\caption{
(a) Top view of CuO$_2$ layer: blue balls are Cu, and red balls are O.  The black ball shows the projected position of the nearest O dopant between the Bi and Sr layers while the black dashed line marks the $a$-axis position of this dopant. 
(b) The AFM correlation $\avg{\hat{N}_{i\uparrow} \hat{N}_{j\downarrow}+\hat{N}_{i\downarrow} \hat{N}_{j\uparrow}}_s$ between nearest-neighbor Cu from the cluster slave-boson calculations (a Cu pair is identified by the position of their linking planar O atom);  the O positions along the $a$-axis are in fractions of $a$.  Any inhomogeneity along the $b$-axis is averaged.  Blue circles show the results from the relaxed $x=0.25$ hole-doped structure which has superlattice modulation (SM).  Red diamonds show data for the symmetrized Wannier model which is spatially uniform. 
(c) The local onsite energy difference $\epsilon_{pd}$ for the DFT NM state.   Each data point shows the averaged $dp$-splitting between an oxygen atom and its two copper neighbors.  Green triangles are for the frozen structure obtained by removing oxygen dopants from the $x=0.25$ structure, and purple stars are for the relaxed (modulated) undoped $x=0$ structure. 
(d) The local CuO bond length between an O atom and its two Cu neighbors averaged over the two bonds. 
}
\label{fig:imhomogeous}
\end{figure}

Due to the atomically flat and clean BiO-terminated surfaces that form on BSCCO upon cleaving of crystals, surface tunneling microscopy and spectroscopy can provide a wealth of spatially-resolved electronic information.  A striking experimental discovery 
\cite{mcelroy2005atomic, andersen2007superconducting} was a significant modulation of the superconducting gap with local structural motifs such as (inferred) dopant positions or supermodulation. This experimental information is of significant theoretical interest because any microscopic model for superconductivity will have a dependence on the local doping level as well as the local structural details: for example, a leading theory for the microscopic pairing mechanism is through AFM spin correlations and fluctuations \cite{taillefer2010scattering, Jin2011} and local doping levels and inter-site hopping parameters (which depend on local bond lengths and angles) help determine the magnetic properties.  

However, it is very difficult to directly interpret the experimental data without theoretical models.  To date, a large number of theoretical works have attempted to describe the situation \cite{foyevtsova2009effect, gastiasoro2018enhancing, nunner2005dopant, maska2007inhomogeneity, khaliullin2010enhanced, litak2009charge, romer2013modulations, johnston2009impact} using model Hamiltonians and perturbative expansions, but there is a lack of consensus on the dominant expansions terms, and a number of key parameters must be fit to experiment.  Separately, a number of these works assume that the electrostatic potential of the dopant plays the main physical role, an assumption whose validity is hard to assess without independent and parameter-free theoretical results.

In this section, we use our DFT+many-body calculations to shed light on some key questions (the methodology is the same as Section \ref{sec:clustersb} and use the same PM state below).  We begin with Fig. \ref{fig:imhomogeous}(a) that illustrates a top view of the CuO$_2$ layer we focus on and marks the position of the (in-plane projection)  of the oxygen dopants (which are at the necking region of the superlattice modulation).  Fig. \ref{fig:imhomogeous}(b) shows the many-body results for the modulation of AFM correlation along the superlattice modulation direction ($a$-axis) for the relaxed $x=0.25$ system.  The correlation strength is larger closer to the dopant oxygen, which is consistent with experimental observations \cite{mcelroy2005atomic, andersen2007superconducting}. For reference, a flat red line marks the AFM correlation from solving a symmetrized Hubbard model with all hopping strengths and onsite energies set to their average value over the unit cell.

Next, we investigate the relation between the AFM correlation and the local onsite energy difference $\epsilon_{dp}$ between the Cu-$d_{x^2-y^2}$ and O-$p_\sigma$ Wannier orbitals (which, as explained above, are found by Wannierization of the NM DFT+U state within a three-band model). Fig. \ref{fig:imhomogeous}(c) shows the local $dp$ splitting $\epsilon_{pd}$, and we see a clear modulation and direct connection to the AFM correlation strength.  The connection is easy to understand: larger $\epsilon_{dp}$ leads to a lower effective hopping between nearest-neighbor Cu sites and reduces Cu-Cu magnetic couplings (see, e.g. Ref.~\cite{foyevtsova2009effect}).  We expect $\epsilon_{pd}$ to be directly connected to the more experimentally relevant ``charge transfer gap'' $\Delta_{pd}$(between the top of the O-$p$ type hole states and Cu-$d$ type electron states in experiment): for a fixed Cu $U$ and near half-filling, one expects $\Delta_{pd}\approx \epsilon_{pd}+U/2$.  In this way, we expect the modulation of $\epsilon_{pd}$ should reflect that of the charge-transfer gap: Fig. \ref{fig:imhomogeous}(c) predicts an $\approx \pm 0.1$ eV modulation of the charge-transfer gap which is consistent with STM measurements \cite{o2022electron}. 
In addition, since our results show that smaller $\epsilon_{pd}$ leads to stronger AFM correlation, within the spin fluctuation paring model we would expect this leads to stronger superconducting pairing which is consistent with experimental observations \cite{wang2023correlating, ruan2016relationship}.  

We now exploit the power of first-principle calculations, namely that they can deliver parameter-free results along with full control of the microscopic structure, to clearly separate the role of doping from that of structural distortions.  We first take the fully relaxed $x=0.25$ structure which has dopants and supermodulation and simply remove the oxygen dopants without allowing any atoms to move (i.e., frozen structure).  We then recalculate the electronic structure and show the resulting $\epsilon_{pd}$ in green in Fig.~\ref{fig:imhomogeous}(c): the main physical effect is that $\epsilon_{pd}$ curve shifts up uniformly in energy upon removal of the dopants without much change of the amplitude of spatial modulation.  Hence, the electrostatic effect and doping profile from the oxygen dopants are in fact quite uniform (i.e., non-local) and do not in themselves create any strong spatial modulation of electronic parameters.  Obviously, this argues against a key assumption of many prior model Hamiltonian works and helps us focus on other relevant sources for the modulation (i.e., structural properties).  The uniform upward shift of $\epsilon_{pd}$ with the removal of hole dopants is easy to understand: the Cu now becomes more negatively charged and thus their onsite energies get raised up in energy.  For reference, we also relax the new $x=0$ structure which retains the superlattice modulation but with weaker amplitude: its resulting $\epsilon_{pd}$ in Fig.~\ref{fig:imhomogeous}(c) is consistently higher than the $x=0.25$ case but is much flatter than the unrelaxed $x=0$ structure.

While there are many potential structural parameters or descriptors one could investigate, we focus on the simplest and most important one for $\epsilon_{pd}$: the Cu-O bond length.  Fig. \ref{fig:imhomogeous}(d)
shows the local Cu-O bond length in the CuO$_2$ plane for all the four systems described above.  In the necking regions of the supermodulation, which is also where the dopants are located, the local Cu-O bonds are longest which translates into small local $\epsilon_{pd}$.  The connection between Cu-O bond length and $\epsilon_{pd}$ is easy to understand, e.g., from classical electrostatics: for larger Cu-O separation, the Madelung electrostatic onsite potential energies are lowered at Cu and raised at O which reduces $\epsilon_{pd}$. The inverse relation of the two is easy to see when comparing comparable curves in Figs. \ref{fig:imhomogeous}(c) and (d).

In short, we have shown quite clearly that local structural modulations are the dominant controllers of the local $dp$ energy splitting and thus the local AFM correlation strength.  In turn, this structural effect helps us understand the cause of variations in the charge-transfer gap and the superconducting gap observed in experiments.  Furthermore, it highlights the point that controlling and engineering the local structure of the superconducting layers is the most straightforward path to engineering the superconductivity of the cuprates.

\section{Outlook}

In conclusion, we have provided a microscopic understanding of the AFM insulating phase of the undoped Bi-2212 system.  Additionally, we have uncovered competing stripe orders in the hole-overdoped system, which offers a paradigmatic approach (specifically for BSCCO) to describe stripe orders using DFT without performing complex many-body calculations.  Spectroscopically, our non-magnetic DFT band structure calculations remarkably reproduce the observed normal state spectral properties.  Furthermore, we have elucidated the structural origin of the ARPES $\pm(\pi,\pm\pi)$ and $\pm(\pi/4,-\pi/4)$ shadow bands in the hole-overdoped system.  Finally, we show how the dopant oxygens and structural modulations vary the charge-transfer gap and spin correlations over real space: in particular, we show that the local structure around the CuO$_2$ plane is of paramount importance in controlling the local AFM correlation strength and charge-transfer gap, while the effect of the electrostatic doping effect itself is quite minimal.  Our work underscores the importance of considering the crystal degrees of freedom, including structural distortions, modulations, and realistic oxygen dopant positions, together with state-of-the-art exchange-correlation functional, for an accurate description of various material properties.  A telegraphic summary is  ``structure is king.''

We believe that our study establishes a robust theoretical foundation for more ``surgical'' structural engineering of cuprates, particularly with regard to manipulating broken translational and rotational symmetries.  Moreover, as our DFT ground state captures the dominant low-energy properties in the normal state of Bi-2212, the Wannierized Hamiltonians extracted from our DFT calculations can provide appropriate parameters for model Hamiltonians in future studies of low-energy phenomena in cuprates.  As an example, we show in our work that the DFT-derived band structure is in excellent agreement with experiments when computed for using non-magnetic phases.  Using many-body calculations, we show that the reason for this agreement is due to the fact that the overdoped material is in a paramagnetic phase with strong dynamical spin fluctuations (which are known to affect superconductivity and the pseudogap \cite{zhu2023spin,moriya2000spin}) but without any static spatial symmetry breaking of the local moments or charge.  These observations also provide guidance to future researchers regarding which DFT predictions can be compared directly to experimental observables and which other predictions require further refinement using many-body correlated electron approaches.

\section{Computational Details}

We use the Vienna ab initio simulation package (VASP) with the projector-augmented wave method \cite{kresse1999ultrasoft}.  A relatively high plane-wave cutoff energy of 500 eV is used.  The generalized-gradient-approximation (GGA) with the semilocal Perdew–Burke–Ernzerhof (PBE) functional \cite{anisimov1991band, perdew1996generalized} is used in all of our calculations.  All calculations are done with collinear spins.  The recent finding of non-collinear spin texture in cuprates \cite{gotlieb2018revealing} is also an interesting topic to study but is beyond the scope of this work.  To avoid the known failure of the local-spin-density approximation (LSDA) and the GGA in reproducing the copper magnetic moment in cuprates \cite{mattheiss1988electronic, hybertsen1988electronic, lin2006raising, foyevtsova2010modulation, song2019visualization, he2008supermodulation, fan2011modulation} due to self-interaction errors (SIE) in the approximate exchange-correlation functionals \cite{perdew1981self}, we add $U=4$ eV for the Cu $3d$ manifold in all our PBE+$U$ calculations following previous theoretical works \cite{yelpo2021electronic, wang2006oxidation, deng2019higher}.  In the supplementary material \cite{supp}, we describe further tests that vary $U$ from 0-4 eV.  At $U=0$ eV, we reproduce the non-magnetic metallic ground states from prior works \cite{lin2006raising, foyevtsova2010modulation, song2019visualization, fan2011modulation}.  Once $U>2$ eV, changing the $U$ value only shifts the energy of the unoccupied high-energy Cu-derived bands \cite{supp} which does not affect our main findings around the Fermi energy.  

More advanced and superior exchange-correlation functionals such as the strongly-constrained-and-appropriately-normed (SCAN) meta-GGA functional \cite{sun2015strongly} can be employed, and they usually require smaller $U$ value (as they inherently better remove self-interaction errors).  We have performed SCAN+U calculations as well (see the supplement \cite{supp}) using $U$ values for prior works for cuprates \cite{long2020evaluating}.  For both undoped and doped Bi-2122, we find that the SCAN+U method \cite{gautam2018evaluating, long2020evaluating} provides very similar results to the PBE+U results \cite{supp}.  Details on optimized lattice structures as well as typical meta-stable structures can also be found in the supplementary materials \cite{supp}.  Given the similarity of the PBE+U and SCAN+U results and considering the increased computational cost of using SCAN+U, most of the results in our paper use the PBE+U approach.

To accelerate the structural relaxation of the large 244-atom oxygen-doped unit cell, we use the Spanish Initiative for Electronic Simulations with Thousands of Atoms (SIESTA) package \cite{soler2002siesta} to approximately relax the structure before doing the final relaxations using VASP.  A DZP basis with EnergyShift of 100 meV and SplitNorm of 0.25 is used in all our SIESTA calculations. 

We substantiate this picture quantitatively by computing the tight-binding Kohn-Sham Hamiltonian on the maximally localized Wannier basis \cite{marzari1997maximally} extracted from our DFT calculations using Wannier90 \cite{pizzi2020wannier90}.

Our slave-boson calculations use the Wannierized tight-binding model from our DFT calculations to construct the Hubbard model.  We perform the single-site slave-boson calculations using the Boson Subsidiary-Solver (BoSS) software \cite{georgescu2021boson}.  The cluster slave-boson calculations follow early work \cite{jin2023bond}. 

\section{Acknowledgements}
We thank Byungmin Sohn and Yu He for their helpful discussions and comments on the manuscript. This work was supported by grant NSF DMR 2237469, NSF ACCESS supercomputing resources via allocation TG-MCA08X007, and supercomputing resources from Yale Center for Research Computing.

\appendix

\section{Three-band versus one-band model}
\label{sec:threeband}
\begin{figure}[t]
\begin{center}
\includegraphics[scale=0.26]{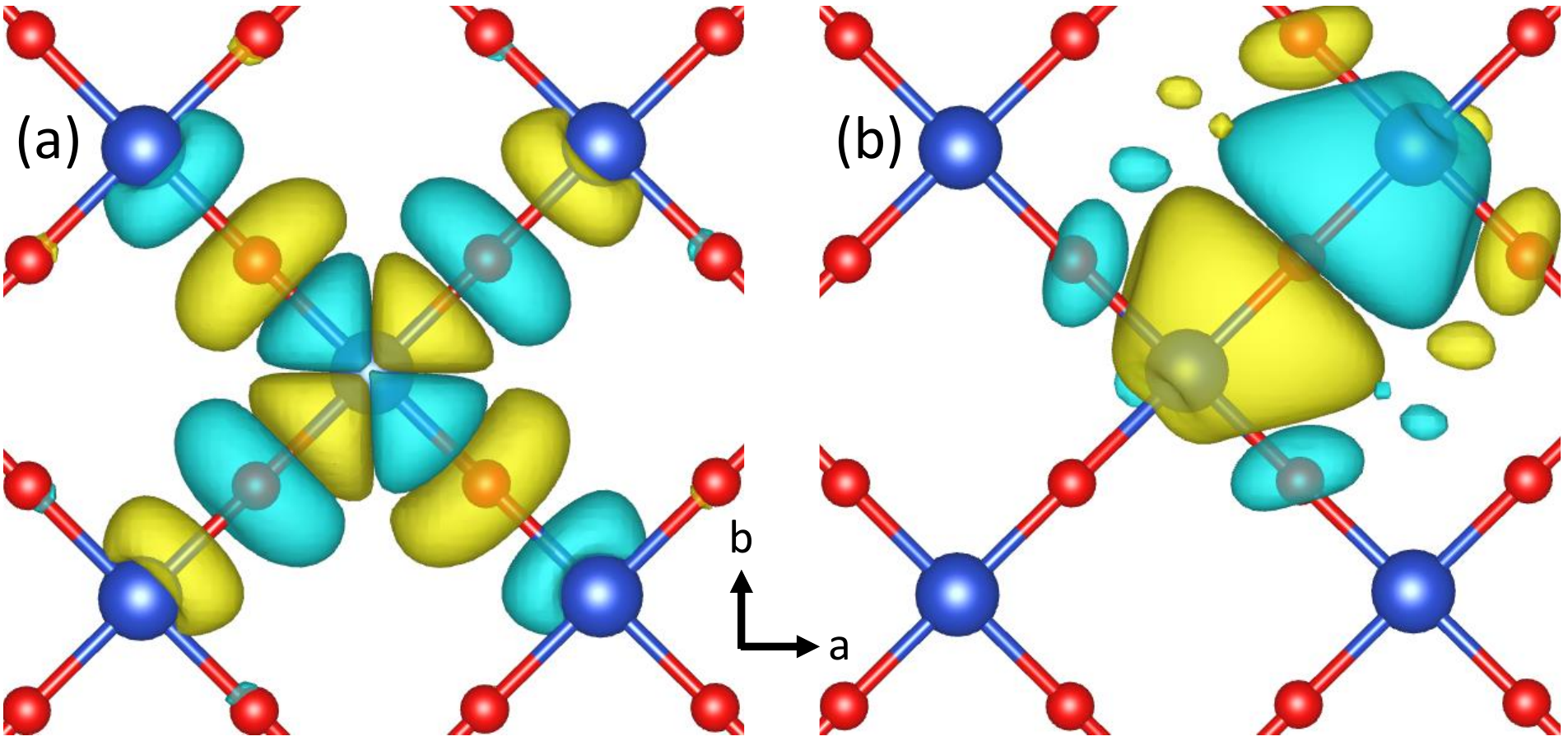}
\end{center}
\caption{
Top view of the maximally-localized Wannier functions' isosurface in real space for the three-band model.  (a) Cu-centered Wannier functions with $d_{x^2-y^2}$ symmetry.  (b) O-centered Wannier functions with $p_x$ symmetry.  The isosurface level is $0.6e$/\AA${^3}$, about 2\% of the maximum value.  Blue large balls are copper atoms, and red small balls are oxygen atoms.  Yellow and blue iso-surfaces represent positive and negative wavefunctions. 
}
\label{fig:wannier_wf}
\end{figure}

Conventionally, the low-energy effective model can also be constructed from the Cu $d_{x^2-y^2}$ orbitals and the O $p_{\sigma}$ ($p_{x/y}$ orbitals pointing toward Cu atoms). This model is often known as the ``three-band model'' or Emery model \cite{emery1987theory}.  In this section, we will compare the three-band model with the one-band model in the main text.  It turns out that they provide equivalent descriptions for the bands around the Fermi level.  Hence, the choice of different effective models does not affect the results in the main text. 

Here we construct a Wannier basis with Wannier functions centered at Cu and O atoms on the CuO$_2$ layers via the maximally localized Wannier function method.  The Cu-centered Wannier functions are constructed to have the same symmetry as $d_{x^2-y^2}$ orbitals, while the O-centered Wannier functions have the $p_\sigma$ symmetry.  Fig. \ref{fig:wannier_wf} shows the isosurface of the Wannier functions in real space, where both types of Wannier functions are well-localized.  Consequently, our effective model contains three Wannier orbitals for each CuO$_2$.  Different from the idealized three-band model in prior studies, our model contains important symmetry-breaking information from structural distortions.  For example, the on-site energies and hopping strengths vary among different atoms and bonds.   

\begin{figure}[t]
\begin{center}
\includegraphics[scale=0.35]{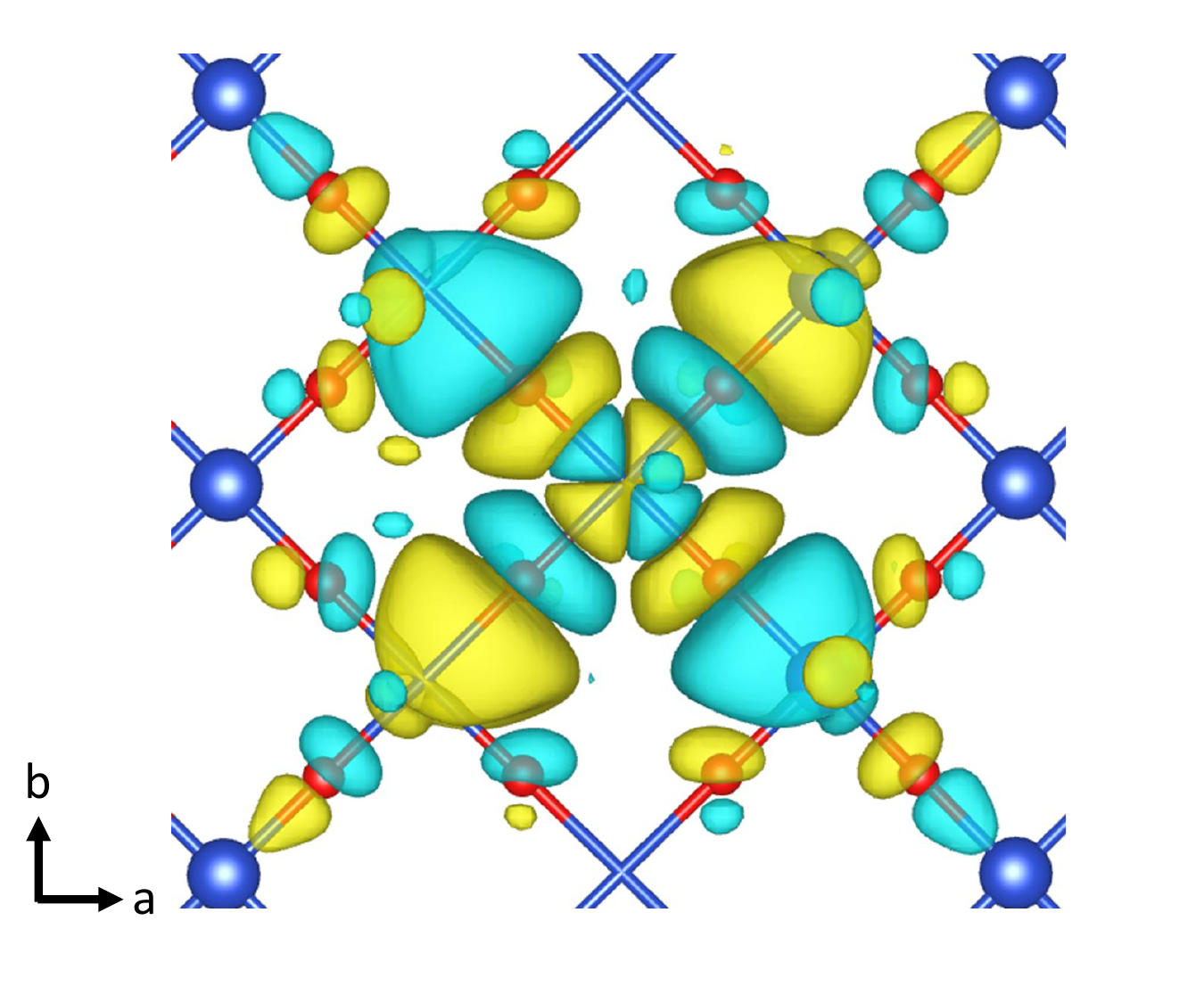}
\end{center}
\caption{
Top view of the maximally-localized Wannier functions' isosurface in real space for the one-band model.  The isosurface level is $0.6e$/\AA$^3$.  Blue large balls are copper atoms, and red small balls are oxygen atoms.  Yellow and blue iso-surfaces represent positive and negative wavefunctions. 
}
\label{fig:wannier_wf_singleband}
\end{figure}
\begin{figure}[t]
\begin{center}
\includegraphics[scale=0.55]{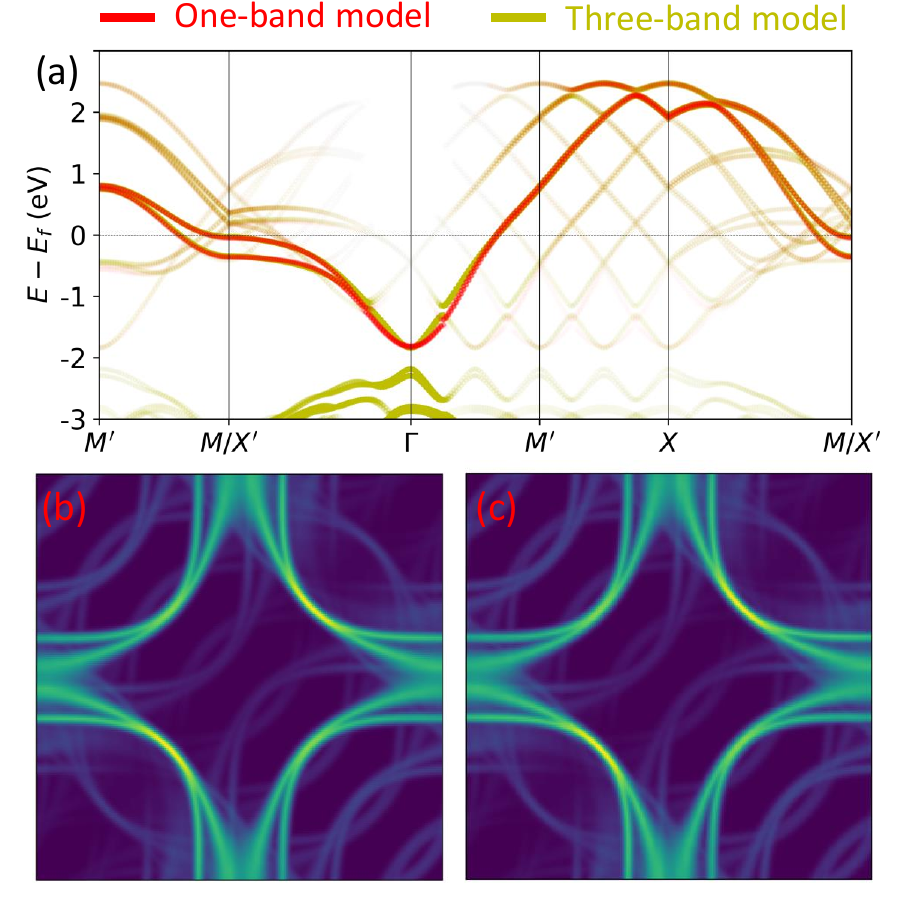}
\end{center}
\caption{
(a) Comparison of electronic structures from one-band and three-band models in the hole-overdoped BSCCO.  The red curves represent results from the one-band model and the yellow curves represent the three-band model.  (b-c) The corresponding Fermi surfaces from one-band and three-band models, separately.  
}
\label{fig:one_three_band}
\end{figure}
As a comparison, the one-band model in the main text is physically a low-energy simplification of the three-band model by focusing on the anti-bonding band of CuO$_2$.  Fig. \ref{fig:wannier_wf_singleband} shows the maximally-localized Wannier function of the one-band model.  It is centered at the Cu atom and comes from the antibonding hybridization of the three-band model Wannier functions in Fig. \ref{fig:wannier_wf}.  

We further compare the electronic structures of the one-band model and the three-band model.  Fig. \ref{fig:one_three_band}(a) shows the unfolded band structures of the 25\% hole-doped BSCCO in the NM state.  These two models give quantitative similar results for the anti-bonding CuO$_2$ bands crossing the Fermi level, i.e. all the bands between $-2eV$ to $2eV$ around the Fermi level.  They almost exactly overlap with each other.  The three-band model shows additional fully occupied bands below $-2eV$ from the Fermi level.  These fully occupied bands are beyond our interest because they have little effect on the Fermi surface.  Fig. \ref{fig:one_three_band}(b) and (c) show the Fermi surfaces from the one-band and three-band
models, separately.  As expected, there is no visual difference between the Fermi surfaces, including the shadow band intensities. 

\begin{figure}[t]
\begin{center}
\includegraphics[scale=0.65]{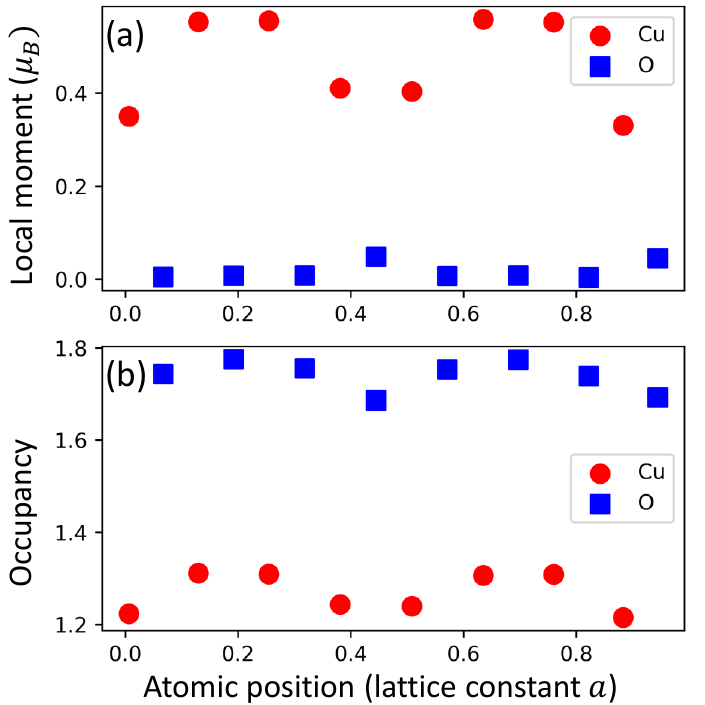}
\end{center}
\caption{
(a) Local moments magnitude and (b) Electron occupancy at different sites along the $a$-axis for the three-band model describing the  stripe order state shown in Fig. \ref{fig:stripe_order}(b). The magnitude of local moments in $\mu_B$ on sites $i$ along $a$-axis defined as $|m_i|\equiv |n_{i\uparrow}-n_{i\downarrow}|\mu_B$, where $n_{i\sigma}$ is the occupancy of the Wannier orbital site.  Red circles and blue squares represent the Wannier orbitals centered at the Cu and O atoms, separately.   
}
\label{fig:stripe_order_3band}
\end{figure}

Finally, we study the local moments and the electron occupancies in the three-band model as a comparison to the one-band model results in Fig. \ref{fig:stripe_order}.  Here we focus on the same stripe order shown in Fig. \ref{fig:stripe_order}(b) and Wannierize it to the three-band model.  Fig. \ref{fig:stripe_order_3band}(a) shows the local moments of different Cu $d_{x^2-y^2}$ and O $p_{\sigma}$ Wannier orbitals in the unit cell.  The dominant local moments are on Cu, showing a consistent wavy pattern as the one-band model results in Fig. \ref{fig:stripe_order}(c).  Fig. \ref{fig:stripe_order_3band}(b) shows the electron occupancies of different Wannier orbitals, where both Cu and O Wannier orbitals show a consistent wavy pattern as the one-band model results in Fig. \ref{fig:stripe_order}(d).  This analysis provides a consistent picture in a prior work on YBCO \cite{zhang2020competing}. 

\section{Stripe order states and their energies in hole-doped system}
\label{app:supp_stripe}
\begin{figure*}
\begin{center}
\includegraphics[scale=0.48]{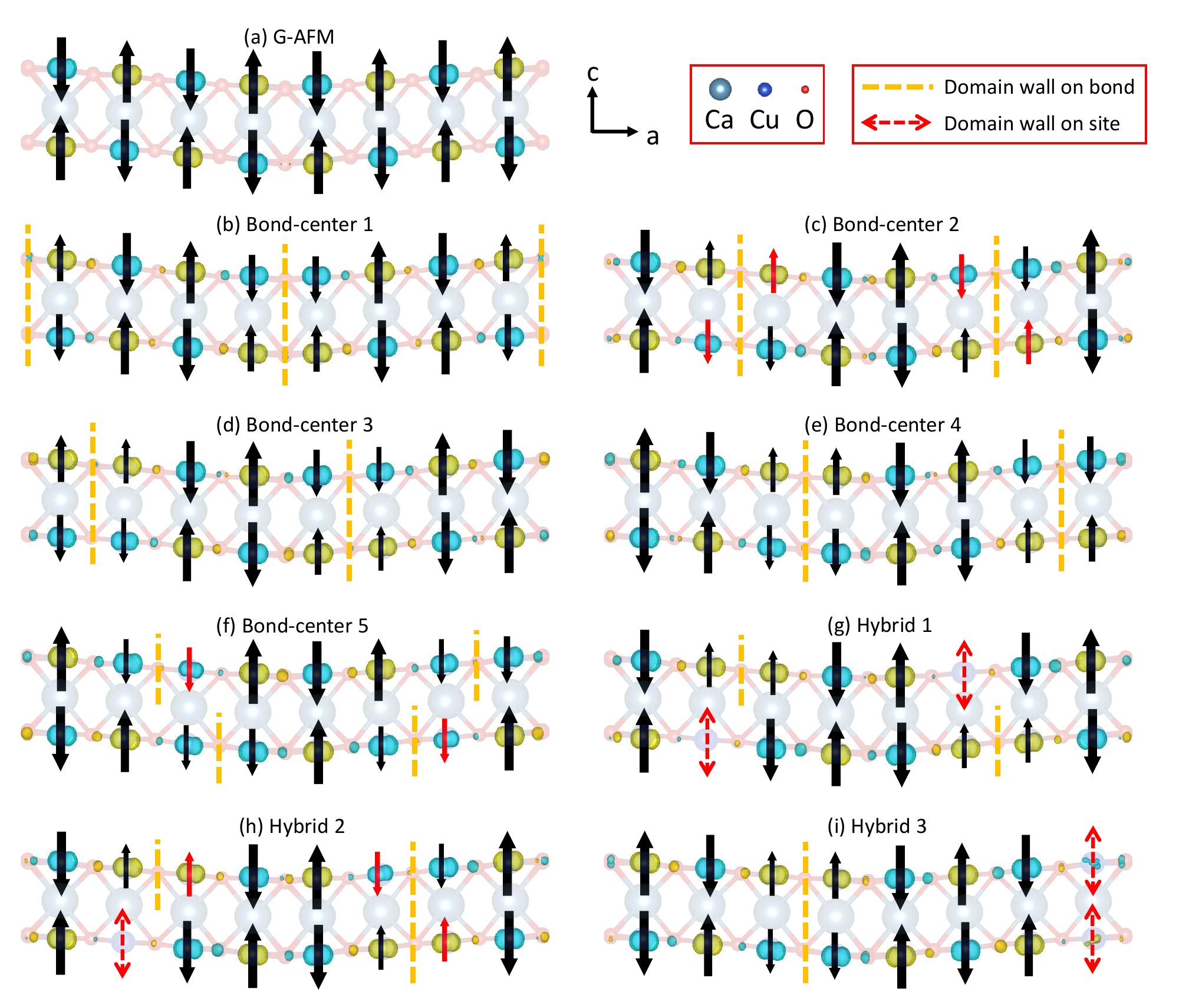}
\end{center}
\caption{
Spin densities of the G-AFM state and stripe order states.  The yellow isosurfaces show spin-up polarization densities, while the blue isosurfaces show spin-down densities.  The arrows illustrate the spin directions for clarity.  The local moments on the Cu atoms are classified into three types:  The big black arrows represent the local moments larger than 0.4$\mu_B$, the red arrows or dashed double arrows represent those smaller than 0.25$\mu_B$, and all other local moments are marked by small black arrows.  The orange dashed lines show the domain walls on Cu-Cu bond, while the red dashed double arrows show the domain walls on Cu sites.  (a) The G-AFM order state.  (b-f) Five typical bond-center stripe orders.  (g-i) Three typical hybrid stripe orders containing domain walls on bonds and sites at the same time.  
}
\label{fig:supp_244_stripe_patterns}
\end{figure*}

\begin{table*} 
\caption{
The local magnetic moments and total energies of the stripe order states.  The total energy of the G-AFM state is set to be the reference energy.  Here we classify the local moments as the ``Big black arrows'', ``Small black arrows'', and ``Red arrows'' according to the illustrations in Fig. \ref{fig:supp_244_stripe_patterns}. 
}
\begin{ruledtabular} 
\begin{tabular} { c c c c c }
Spin structure & \multicolumn{3}{c}{Local moment amplitudes ($\mu_B$/Cu)} & Energy\\
~ & Big black arrows & Small black arrows & Red arrows & (meV/Cu)\\
\hline
G-AFM & 0.42$\sim$0.46 & - & - & 0 \\
Bond-center 1  & 0.41$\sim$0.49 & 0.34$\sim$0.36 & - & 1.9 \\
Bond-center 2  & 0.42$\sim$0.52 & 0.32$\sim$0.33 & 0.23$\sim$0.25 & -0.3 \\
Bond-center 3  & 0.42$\sim$0.51 & 0.30$\sim$0.38 & - & 1.5 \\
Bond-center 4  & 0.43$\sim$0.51 & 0.33$\sim$0.37 & - & 2.1 \\
Bond-center 5  & 0.43$\sim$0.52 & 0.30$\sim$0.32 & 0.16$\sim$0.18 & 1.4 \\
Hybrid 1  & 0.42$\sim$0.52 & 0.37$\sim$0.39 & $<$0.03 & 0.5 \\
Hybrid 2  & 0.42$\sim$0.52 & 0.34$\sim$0.39 & 0$\sim$0.19 & -0.1 \\
Hybrid 3  & 0.40$\sim$0.51 & 0.28$\sim$0.29 & $<$0.04 & 2.7 
\end{tabular} 
\end{ruledtabular} 
\label{tab:supp_stripes}
\end{table*}

 
As discussed in the main text, we have successfully identified many stripe-ordered states that are almost degenerate with the G-AFM state in our DFT calculations.  We conclude that there is likely strong spin fluctuations in this overdoped system.  Such spin fluctuations in cuprates are also detected by experiments \cite{zaanen1996dynamical, lyons1988dynamics, lyons1988spin, brom2003magnetic, zhu2023spin}.  In this section, we provide more details about these competing ordered states.

Table \ref{tab:supp_stripes} shows the local moments of eight distinct stripe order states and their total energies compared to the G-AFM state.  We identify several bond-centered stripe orders, where all the domain walls are located at the O atomic sites mediating the Cu-Cu bonds.  Figure. \ref{fig:supp_244_stripe_patterns}(b-f) show the spin structures of these bond-centered stripe orders, where the domain walls marked by the dashed lines can appear at each Cu-Cu bond along $a$-axis.  While the domain walls are separated by four Cu atoms in Fig. \ref{fig:supp_244_stripe_patterns}(b-e), it is also possible to have three or five Cu atoms between domain walls as shown in Fig. \ref{fig:supp_244_stripe_patterns}(f).  

In addition to the bond-centered stripe orders, a previous DFT study on YBCO has also identified site-centered stripe orders \cite{zhang2020competing}, where all domain walls are at Cu atomic sites.  However, such site-centered stripe orders are unstable in Bi-2212 according to our calculations.  Instead, we find several ``hybrid'' stripe orders, where domain walls appear both at the Cu-Cu bonds and the Cu atomic sites as illustrated in Fig. \ref{fig:supp_244_stripe_patterns}(g-i).  The red dashed double arrows show the domain walls on Cu atomic sites, where the local moments are almost zero.  Interestingly, we find several Cu sites with very small local moments marked by the red arrows, which are very common in the stripe patterns such as those in Fig. \ref{fig:supp_244_stripe_patterns}(c), (f), and (h).  The physical origin of this phenomenon is likely due to the superlattice modulation distortion which modulates the local environments of the Cu atoms.

In principle, there are many more possible configurations of bond-centered stripe orders, because one can take combinations of the existing stripe patterns for the upper and lower CuO$_2$ layers.  We have tested 9 additional bond-centered patterns by taking these combinations and found energies of 1.65$\sim$3.40 meV/Cu above the G-AFM state.  These stripe-order states, constructed by such combinations, generally show higher total energies than the existing stripe-order states (but all of them are very close in energy).  All these calculations indicate the energetic tendency for the existence of strong spin fluctuations in this system: as we show in the main text, one has to go beyond DFT to take account of the spin fluctuation effect to make quantitative predictions of ARPES spectra.

\bibliography{v1}
\end{document}